\documentclass[amsmath,amssymb,aps,pra,twocolumn,showpacs,10pt,floatfix]{revtex4-1}%,superscriptaddress
\usepackage{graphicx}% Include figure files
\usepackage{dcolumn}% Align table columns on decimal point
\usepackage{bm}% bold math
\usepackage{dsfont}
\usepackage[utf8]{inputenc}
\usepackage{hyperref}

\begin{document}

\newcommand{\ave}[1]{\left\langle #1\right\rangle}
\newcommand{\wv}[3]{{}_{#1}\ave{#2}_{#3}}
\def\Tr{\mathrm{Tr}}
\def\tr{\mathrm{tr}}
\def\re{\mathrm{Re}}
\def\im{\mathrm{Im}}
\newcommand{\bra}[1]{\left<{#1}\right|}
\newcommand{\ket}[1]{\left|{#1}\right>}
\newcommand{\kett}[1]{|{#1}\rangle}
\newcommand{\braket}[2]{\left<\left.{#1}\right|{#2}\right>}
\newcommand{\expt}[3]{\left<{#1}\left|{#2}\right|{#3}\right>}
\def\pdag{{\phantom{\dagger}}}
\newcommand{\rs}{\rm \scriptscriptstyle}
\def\alphare{\alpha_{\re}}
\def\alphaim{\alpha_{\im}}
\def\betare{\beta_{\re}}
\def\betaim{\beta_{\im}}
\newcommand\mdoubleplus{\mathbin{+\mkern-10mu+}}

\title{Distinguishing phases using the dynamical response of driven-dissipative light-matter systems}
\date{\today}

\author{M. Soriente}
\affiliation{Institute for Theoretical Physics, ETH Zurich, 8093 Z{\"u}rich, Switzerland}
\author{R. Chitra}
\affiliation{Institute for Theoretical Physics, ETH Zurich, 8093 Z{\"u}rich, Switzerland}
\author{O. Zilberberg}%
\affiliation{Institute for Theoretical Physics, ETH Zurich, 8093 Z{\"u}rich, Switzerland}
%\pacs{42.50.Pq, 05.30.Rt, 32.80.Qk, 42.65.Yj}

\begin{abstract}
We present a peculiar transition triggered by infinitesimal dissipation in the interpolating Dicke-Tavis-Cummings model. The model describes a ubiquitous light-matter setting using a collection of two-level systems interacting with quantum light trapped in an optical cavity. In a previous work [Phys. Rev. Lett. \textbf{120}, 183603 (2018)], dissipation was shown to extend a normal phase (dark state) into new regions of the model's parameter space. Harnessing Keldysh's action formalism to compute the response function of the light, we show that the normal phase does not merely spread but encompasses a transition between the old and the dissipation-stabilized regimes of the normal phase. This transition, however, solely manifests in the dynamical fluctuations atop the empty cavity, through stabilization of an excited state of the closed system. Consequently, we reveal that the fluctuations flip from being {\it particlelike} to {\it holelike} across this transition. This inversion is also accompanied by the behavior of the Liouvillian eigenvalues akin to exceptional points. Our work forges the way to discovering transitions in a wide variety of driven-dissipative systems and is highly pertinent for current experiments.
\end{abstract}

\maketitle
\section{Introduction}
Driven-dissipative many-body systems provide a rich arena to explore many-body out-of-equilibrium phases of matter~\cite{Petruccione}. Within this realm, light-matter systems are prototypical representatives, as they are unavoidably coupled to a bath and in general also subject to external drives~\cite{Carusotto_2013,Ritsch_2013,Baumann_2010,Klinder_2015,Zhiqiang_2017,Morales_2019,Chiacchio_2018_2,Reiter_2018,Lambert_2019}. Light-matter systems additionally offer a high-degree of experimental control, thus enabling tunable realizations of complex configurations and the concomitant observation of exotic phenomena, e.g., polariton condensates~\cite{Kasprzak_2006}, superradiant lasing~\cite{Bohnet_2012,Kirton_2018,Laske_2019}, and supersolid formation~\cite{Leonard_2017_n,Mivehvar_2018}.

Dissipation channels play a crucial role in dictating the behavior of light-matter systems. Most commonly, they act as a sink to the energy provided by the various drives, and stabilize the system's dynamics into a ground state of the corresponding closed system. Interestingly, dissipation can also give rise to new dissipative phase-transitions~\cite{Soriente_2018,Dogra_2019,Chiacchio_2019,Buca_2019,Fitzpatrick_2017,Fink_2018}, complex dynamics~\cite{Konya_2018,Puebla_2019}, the emergence of new universality classes~\cite{Zamora_2017,Zhang_2019,Xu_2019}, and non-Hermitian phases governed by exceptional points~\cite{Heiss_2012,Ozdemir_2019,Hanai_2019}. Crucial to this work is the exploration of dissipation as a tool to generate new dynamical phases inaccessible in closed systems.

The open interpolating Dicke-Tavis-Cummings (IDTC) model is a paradigmatic driven-dissipative model that provides an ideal example of the dramatic impact of dissipation on many-body phenomena~\cite{Soriente_2018,Bhaseen_2012}. It effectively describes the physics of a variety of light-matter systems~\cite{Baksic_2014,Kirton_2019}, and in a recent work~\cite{Soriente_2018}, we have shown that dissipation profoundly alters the phase diagram of this model. Specifically, dissipation stabilizes and extends the so-called normal phase (NP) into a new parameter regime, as well as leads to coexistence of phases. Despite the completely different setting in the new region, no static phase transition links the original NP and the dissipation-generated NP.

%%%%%%%%%%%%%%%%%%%%%%%%%%%%%%%%%%%%%%%%%%%%%%%%%%%%%%%%%%%%%%%%%
\begin{figure}[t!]
 \includegraphics[width=\columnwidth]{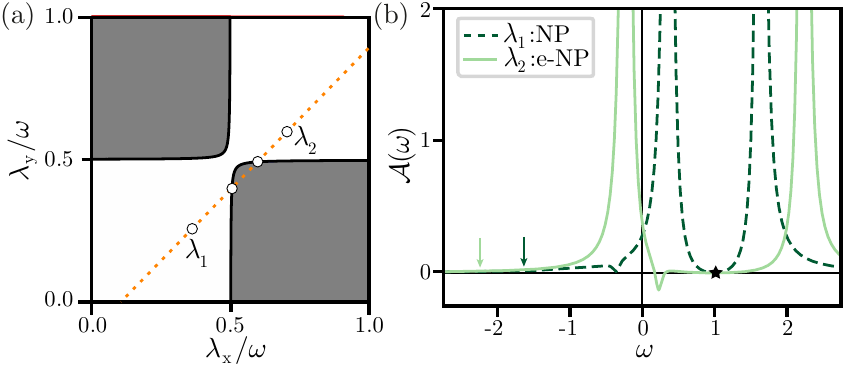}
 \caption{Dissipative interpolating Dicke-Tavis-Cummings model~\cite{Soriente_2018}, cf.~Eq.~\eqref{eq:IDTC}. (a) Steady-state phase diagram: normal phases (white) are stable in the bottom-left (NP) and upper-right (e-NP) quadrants. The orange dotted cut passes through the NP and e-NP phases via the superradiant one (gray region) and is used to show the eigenvalue behavior, see Fig.~\ref{fig:closed_sys}. (b) Cavity dynamical response function over the NP (dashed dark-green) and e-NP (light-green), evaluated at $\lambda_{1,2}$ in (a), respectively. The response always presents four peaks appearing at paired frequencies (arrows indicate unresolved small peaks). Remarkably, there is a soft-mode peak inversion between the two NPs. The non-Lorentzian shape of the peaks and the zero-response point (marked with $\star$) are attributed to a Fano resonance~\cite{Joe_2006}. In all plots, $\omega_c=\omega_a,\kappa/\omega_c= 0.1,\lambda_1=(0.35,0.29),\lambda_2=(0.65,0.59)$.}
 \label{fig:open_sys}
\end{figure}
%%%%%%%%%%%%%%%%%%%%%%%%%%%%%%%%%%%%%%%%%%%%%%%%%%%%%%%%%%%%%%%%%

In this work, we show that a new kind of transition, where dynamical correlation functions act as order parameters, distinguishes between the two NP phases. We analyze this open system physics using a combination of analytical methods, including Keldysh action formalism~\cite{kamenev,DallaTorre_2013,Sieberer_2016}, normal mode symplectic structures~\cite{Bogolyubov_1947,Valatin_1958,Xiao_2009}, as well as third quantization~\cite{Prosen_2008,Prosen_2010}. We identify the new dissipation-stabilized NP as an excited state (dubbed e-NP) in the corresponding closed system, i.e., the dissipation incoherently drives a population inversion in the system. As the e-NP corresponds to an empty cavity, this scenario resembles dark lasing, i.e., the system absorbs photons in order to decay to the superradiant ground state. Concurrently, the e-NP exhibits a negative frequency instability~\cite{Kohler_2018} that manifests as an inversion in the soft-mode fluctuations, i.e., the original NP presents standard fluctuations while the e-NP has anomalous ones. We reveal that the former corresponds to particle-dominated fluctuations, whereas the inversion makes the e-NP exhibits holelike fluctuations. Furthermore, using third quantization~\cite{Prosen_2010}, we show that this many-body dynamical behavior is reminiscent of the physics of exceptional points. Our results can explain the dynamics of a variety of driven-dissipative systems~\cite{Zhiqiang_2017,Dogra_2019,Morales_2019,Lambert_2019}.

\section{The model}
We consider a leaky bosonic cavity mode coupled to $N$ two-level systems~\cite{Soriente_2018,Baksic_2014}, described by the master equation $\frac{d \rho_{\rm sys}}{dt} = - \frac{i}{\hbar} [H(t), \rho_{\rm sys}] + \mathcal{L}[\rho_{\rm sys}]$, where
\begin{align}
 \label{eq:IDTC}
 H & = \hbar\omega_c a^\dagger a + \hbar\omega_a S_z \notag \\
 & + {\frac{2\hbar\lambda_x}{\sqrt{N}}} S_x (a + a^\dagger)+ {\frac{2\hbar\lambda_y}{\sqrt{N}}} i S_y (a - a^\dagger)\,,
\end{align}
is the Hamiltonian of the interpolating IDTC model with $\omega_c$ and $a^\dagger$ the cavity's frequency and creation operator, $\omega_a$ and $S_\alpha = \sum_{j=1}^N \sigma_\alpha^j$ the two-level spacing and collective spin operators describing $N$ identical two-level systems, where $\alpha=x,y,z$ and $\sigma_{\alpha}^{j}$ are Pauli-spin operators. The collective spin couples to both quadratures of the cavity field with couplings $\lambda_x$ and $\lambda_y$, allowing for the interpolation between the Dicke ($\lambda_x\neq\lambda_y$)~\cite{Dicke_1954} and Tavis-Cummings ($\lambda_x=\lambda_y$)~\cite{Tavis_1968} models. In both models, above the critical coupling $\lambda_c = \sqrt{\omega_c \omega_a}/2$, the system transitions from a NP, where the cavity is empty and all two-level spins are oriented along the $z$ axis, to a superradiant phase (SP), where the cavity features a finite mean population and the two-level spins acquire a finite mean magnetization along the $x$ and/or $y$ axes.

The Lindblad dissipator $\mathcal{L}[\rho] = \kappa[ 2 a \rho_{\rm sys} a^\dagger - \{ a^\dagger a , \rho_{\rm sys}\}]$ describes photon loss with rate $\kappa$. In Refs.~\cite{Soriente_2018,Bhaseen_2012} it was shown that the closed system \eqref{eq:IDTC} phase diagram is profoundly altered by dissipation. The most intriguing dissipation-induced feature is the appearance of a NP as the steady-state above criticality~\cite{Soriente_2018}. We denote the latter normal phase, excited-NP or e-NP, as we will reveal below that it is an excited state of the closed system. In what follows, we will focus our study on these two seemingly identical NPs and show through the dynamical response function that their time-dependent fluctuations are intrinsically different. Since spin decay is not expected to qualitatively change the phase diagram we do not consider it explicitly in this work~\cite{Kirton_2017}.

The quantum fluctuation Hamiltonian is obtained by employing Holstein-Primakoff's representation for the spins, $S_+= b^\dagger \sqrt{N - b^\dagger b}$ and $S_z= -\frac{N}{2} + b^\dagger b$, and expanding around the respective mean-field parameters, $ a = \alpha\sqrt{N} + c$ and $b = \beta\sqrt{N} + d$, where $\alpha$ and $\beta$ are complex numbers describing the bosonic coherent states~\cite{Emary_2003,Dimer_2007}. This leads to a typical fluctuation Hamiltonian describing two coupled bosonic fluctuation modes of the form:
\begin{align}
\label{eq:fluc_hamiltonian}
  H_{\rm fl} = & \hbar\omega_c c^\dagger c + \hbar\left(\omega_a + \delta \bar{\omega}_1 \right) d^\dagger d \\
      & +\left(\bar{\lambda}_1 c d^\dagger + \bar{\lambda}_2 c^\dagger d^\dagger + \frac{\delta \bar{\omega}_2}{2} d^2 + \text{H.c.}\right)\,, \notag
\end{align}
where the coefficients $\delta \bar{\omega}_1, \delta \bar{\omega}_2, \bar{\lambda}_1, \bar{\lambda}_2$ depend on the mean-field solution, $\alpha$ and $\beta$~\cite{Soriente_2018} (see Appendix~\ref{appendix_HP} for more details). For both NPs, the mean-field solutions are $\alpha=\beta=0$, and equal-time fluctuations also indicate that the NP and e-NP are equivalent. This leads to the ostensible conclusion that one can continuously deform from one to the other without a transition~\cite{Soriente_2018}.

\section{Dynamical response}
To challenge this premise, we consider here dynamical observables. In particular, we focus on the cavity response function $\mathcal{A}(\omega) = -2\im{[G^R(\omega)]}$, where the retarded Green's function is defined as $G^R(t-t')=-i\theta(t-t')\langle[a(t), a^\dagger(t')]\rangle$. Using Keldysh's action formalism~\cite{DallaTorre_2013,Sieberer_2016}, we calculate the time-dependent response of the dissipative IDTC. The response manifests a fundamental difference between the two NPs, thus unveiling a new kind of transitions in the dynamical response of this system.

The frequency-domain Keldysh action for the fluctuations of the open IDTC model is described by two bosonic modes [cf., Eq.~\eqref{eq:fluc_hamiltonian}], and takes the generic form
\begin{equation}
 \label{eq:Keldyshaction}
 S = \frac{1}{2}\int_\omega \; \Phi^\dagger(\omega)\begin{pmatrix} 0 & {\left[G_{4\times 4}^A\right]}^{-1} \\
             {\left[G_{4\times 4}^R\right]}^{-1} & {D_{4\times 4}^K} \end{pmatrix} \Phi(\omega)\,, %\nonumber
\end{equation}
where $\Phi(\omega) = [\Phi_{\rm cl}(\omega),\,\Phi_{\rm qu}(\omega)]$ is the cavity-spin Nambu 8-spinor composed of the concatenated classical and quantum 4-spinors embodying the classical and quantum Keldysh contour superpositions: $\Phi_{i}(\omega)=\begin{pmatrix} c_{i}(\omega), & c_{i}^*(-\omega), & d_{i}(\omega), & d_{i}^*(-\omega)\end{pmatrix}^T$ with $i={\rm cl},{\rm qu}$, respectively. The corresponding cavity and spin-fluctuation operators are denoted by $c_i$ and $d_i$. The matrices $G_{4\times 4}^A,G_{4\times 4}^R,$ and $D_{4\times 4}^K$ are the advanced, retarded, and inverse Keldysh Green's functions, respectively. Note that we treat dissipation in the aforementioned Lindblad form. Correspondingly, the Green's function matrices are determined by the underlying steady state mean-field solutions of the open system. Since we focus on the two NP phases, in the following, we consider the mean-field solutions, $\alpha=0$ and $\beta=0$~\cite{Soriente_2018} (see Appendix~\ref{appendix_MF}).

%%%%%%%%%%%%%%%%%%%%%%%%%%%%%%%%%%%%%%%%%%%%%%%%%%%%%%%%%%%%%%%%%
\begin{figure}[t!]
 \includegraphics[width=\columnwidth]{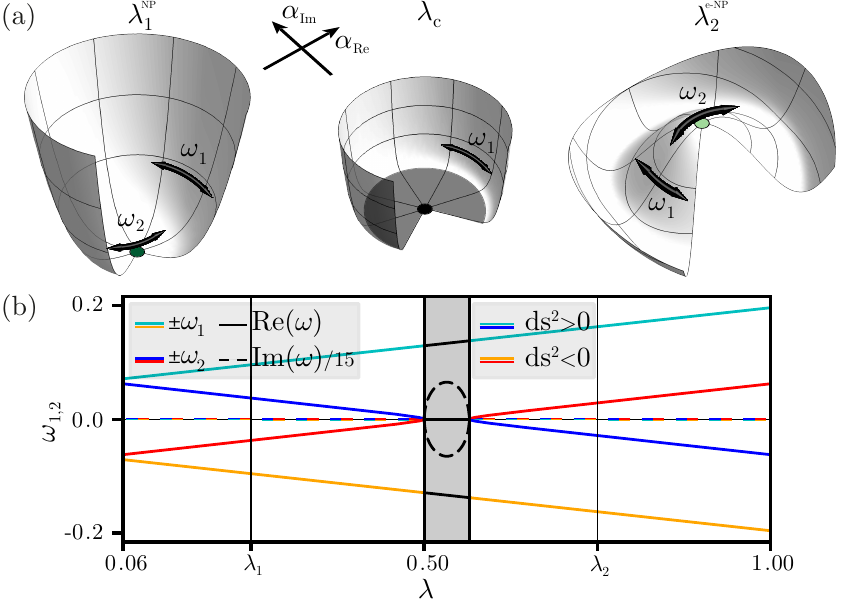}
 \caption{Closed IDTC, cf.~Eq.~\eqref{eq:IDTC}. (a) Mean-field energy landscape as a function of the real and imaginary parts of the cavity field $\alpha=\ave{a}$ at representative points of the parameter space, cf.~Fig.~\ref{fig:open_sys}(a). The fluctuation eigenfrequencies $\omega_1$ and $ \omega_2$ are schematically identified as the azimuthal and radial frequencies of fluctuations around the NP, respectively. Across the NP $\to$ e-NP transition, $\omega_2$ changes sign and is ill-defined at the transition, i.e., at $\lambda_c$. (b) Real (solid) and imaginary (dashed) parts of the fluctuations eigenfrequencies [cf.~Eqs. \eqref{eq:fluc_hamiltonian} and \eqref{eq:NMtransf}] along the orange dotted cut line in Fig.~\ref{fig:open_sys}(a). The eigenfrequencies $\pm\omega_1$ (cyan, orange lines) do not show qualitative changes. The eigenfrequencies $\pm\omega_2$ (blue, red lines) instead become imaginary at $\lambda_c$, signaling the transition NP$\to$SP (shaded region) followed by a return to  real values when $\lambda_x,\lambda_y > \lambda_c$, marking SP$\to$e-NP. Blue and cyan lines are associated with particlelike fluctuations ($ds^2>0$) [cf.~Eq.~\eqref{eq:symplectic_norm}]. Red and orange lines are associated with holelike ones ($ds^2<0$).}
 \label{fig:closed_sys}
\end{figure}
%%%%%%%%%%%%%%%%%%%%%%%%%%%%%%%%%%%%%%%%%%%%%%%%%%%%%%%%%%%%%%%%%

Integrating out the spin fluctuation operators, we obtain a pure cavity four-component field Keldysh action. The resulting retarded Green's function determines the cavity response function $\mathcal{A}\left(\omega\right)$ in the open system setting, see Fig.~\ref{fig:open_sys}(b) (see the detailed derivation in Appendix~\ref{appendix_KA}). In the standard NP, the response is conventional: two positive-amplitude peaks at positive frequencies correspond to two polariton excitations, balanced by as many negative peaks at negative frequencies. The latter is an outcome of the bosonic statistics of the cavity via the normalization condition $\int_\omega \mathcal{A}(\omega) = \langle [a,a^\dagger]\rangle = 1$. On the other hand, the response displays a striking qualitative change as one crosses over into the e-NP: a peak inversion of the softer polaritonic mode occurs. This signature hints at a fundamental difference between the two normal phases. Note that the response also features a Fano resonance~\cite{Fano_1961,Joe_2006}, resulting from the interplay between the two discrete bosonic modes (scattering channels), and the continuous cavity dissipation (decay channel). The superradiant phase presents a conventional response and is therefore not part of our study (see Appendix~\ref{appendix_SR}).

\section{Closed system interlude}
To understand the origin and significance of the peak inversion, it is useful to investigate the mean-field energy of the closed system~\eqref{eq:IDTC}, see Fig.~\ref{fig:closed_sys}(a). It varies from a paraboloid shape with a single minimum (NP) to a distorted ``sombrero-hat'' potential with two distinct minima, indicating the broken $\mathds{Z}_2$ symmetry leading to superradiance (SP)~\cite{Soriente_2018,Baksic_2014}. We remark that the NP is not the ground state in all the symmetry-broken regimes, as it corresponds to a local maximum at the top of the hat-like potentials. Interestingly, however, only in the symmetry-broken regimes, where both $\lambda_x,\lambda_y\ge\lambda_{c}$, the NP manifests as a valid physical mean-field excited state of the system~\cite{Soriente_2018,im_freq}. Nevertheless, it is unstable in the closed system due to a ``negative-frequency instability''~\cite{Kohler_2018} induced by fluctuations, i.e., it decays to the ground state, which is the SP. As we will show below, the negative frequency entails an inversion between particle-dominated fluctuations in the standard NP to hole-dominated fluctuations in the e-NP.

% . Along the $U(1)$ symmetry breaking line, the potential is a minimum along a continuous equipotential circle

To obtain the cavity response of the closed system, we rewrite Eq.~\eqref{eq:fluc_hamiltonian} in terms of its normal modes using the Bogolyubov matrix $T$,
\begin{equation}
 \label{eq:NMtransf}
 \phi = T \psi\,,
\end{equation}
where $\phi = \begin{pmatrix} c & d & c^\dagger & d^\dagger \end{pmatrix}^T$, and $\psi = \begin{pmatrix} A_1 & A_2 & A_1^\dagger & A_2^\dagger \end{pmatrix}^T$ is the normal-mode operator vector~\cite{Prosen_2008,Prosen_2010}. Two key ingredients define this diagonalization procedure: (i) the normal mode eigenfrequencies, $\pm\omega_i$ with $i=1,2$, that determine the phase boundaries when they become complex [cf.~Fig.~\ref{fig:closed_sys}(b)]~\cite{im_freq}, and (ii) the norm associated with the corresponding orthonormal eigenvectors  
\begin{equation}
\label{eq:symplectic_norm}
 ds^2_{i\sigma}\equiv \sum_{j=\{c,d\},\sigma'=\pm}\sigma' |t_{j}^{\sigma'}(\omega_{i\sigma})|^2\,,
\end{equation}
where $\omega_{i\sigma} \equiv \sigma\omega_i$ with $\sigma=\pm$ and $i =1,2$. The sum runs over column elements of the matrix $T$, i.e., $t_{j}^{\sigma'}(\omega_{i\sigma})$, that are indexed according to $j$ and $\sigma'$ (see the detailed derivation in Appendix~\ref{appendix_NM}). We dub Eq.~\eqref{eq:symplectic_norm} \textit{symplectic norm}, reminiscent of an $AdS(2,1)$ norm in symplectic space with particle (hole) entries mapped to space (time).
In Fig.~\ref{fig:closed_sys}(b), we plot the eigenfrequencies and their symplectic norms as a function of the couplings along a trajectory in parameter space that traverses both the NP and e-NP [cf., Fig.~\ref{fig:open_sys}(a)]. The eigenmodes $\pm\omega_1$ show no significant feature, while $\pm\omega_2$, becomes imaginary (nonphysical) when $\lambda_{x,y}\to\lambda_c$, i.e., upon a crossover to the SP. Crucial to our work, both normal mode frequencies are real (physical) in the e-NP. The eigenmodes $\pm\omega_2$, however, flip their norms (particle-to-hole inversion) along the transition from the NP to the e-NP [similar to going beyond the horizon in $AdS(2,1)$], implying that the NP and e-NP are dynamically distinct phases. Schematically, the normal modes 1 and 2 can be visualized as azimuthal and radial fluctuations in the NP mean-field landscape, see Fig.~\ref{fig:closed_sys}(a). The negative radial frequency in the e-NP is consistent with the interpretation of the e-NP as an excited state at the top of the sombrero hat potential.

%%%%%%%%%%%%%%%%%%%%%%%%%%%%%%%%%%%%%%%%%%%%%%%%%%%%%%%%%%%%%%%%%
\begin{figure*}[t!]
 \includegraphics[width=\textwidth]{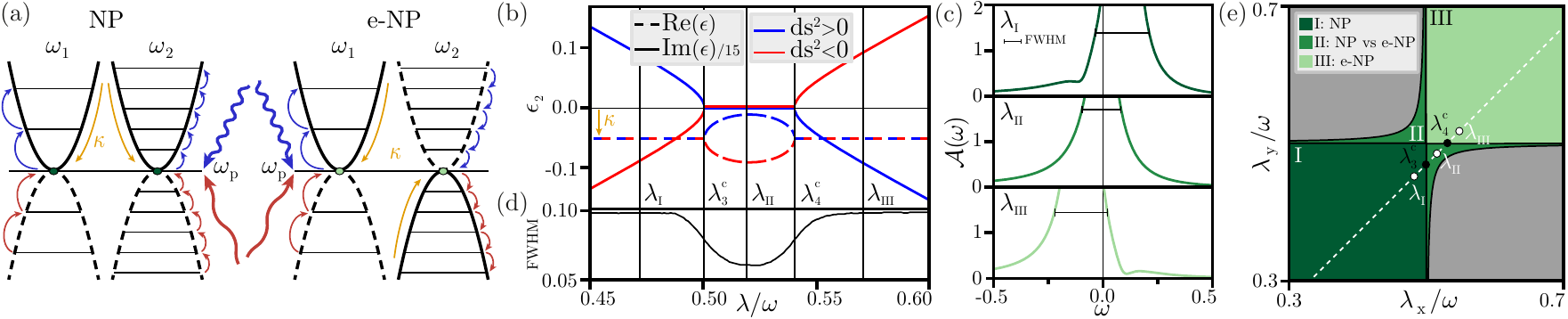}
 \caption{Dark laser and its response. (a) Schematic energy potentials of particle- and holelike fluctuations of the normal modes in the NP (left) and e-NP (right)[cf.~Eqs.~\eqref{eq:fluc_hamiltonian},~\eqref{eq:NMtransf}, and~\eqref{eq:symplectic_norm}]. Blue (red) wiggly lines refer to blue (red) detuned response processes. Solid (dashed) parabolic potentials with companying ladders of levels indicate normal-mode fluctuations around NP with a positive (negative) norm [cf.~Eq.~\eqref{eq:symplectic_norm}]. Solid arrows mark the associated process, i.e., upward blue (downward red) correspond to creation of positive (negative) excitations while downward blue (upward red) to annihilation of such excitations. Orange arrows mark the effect of dissipation in reducing the populations of the modes. (b) Top: Real (dashed lines) and imaginary (solid lines) parts of the soft-mode Liouvillian eigenvalues. Bottom: The extracted FWHM of the positive peak. Parameters along the dashed (white) line in (d). (c) Calculated dynamical response along the NP$\to$e-NP transition. The positive and negative soft-mode peaks invert alongside a region where they overlap. (d) Dynamical phase diagram of the dissipative IDTC for $\kappa/\omega= 0.1$. Region I,III correspond to the NP, e-NP, respectively. Region II is the region where the imaginary parts of the soft mode coalesce to zero.}
 \label{fig:open_final}
\end{figure*}
%%%%%%%%%%%%%%%%%%%%%%%%%%%%%%%%%%%%%%%%%%%%%%%%%%%%%%%%%%%%%%%%%

The closed system cavity response can be straightforwardly evaluated in the normal mode basis (more details in Appendix~\ref{appendix_RF}):
\begin{equation}
 \label{eq:NMresponse}
 \mathcal{A}(\omega)=\sum_{i,\sigma} \frac{|t_{c}^{+}(\omega_{i\sigma})|^2}{ds^2_{i \sigma}} \delta\big(\omega - \omega_{i\sigma}\big)\,.
\end{equation}
As in the open system [cf.~Fig.~\ref{fig:open_sys}(b)], the NP response is characterized by positive (negative) weights at frequencies $\omega_1,\omega_2$ ($- \omega_1,-\omega_2$). Importantly, due to the aforementioned soft-mode particle-hole inversion, the e-NP response is characterized by positive weights at frequencies $( \omega_1,-\omega_2)$ and negative weights at $(- \omega_1,\omega_2)$. This is a key observation of our work, namely, the symplectic norm associated with a normal-mode frequency determines whether the response at that frequency is positive or negative. Modes with a positive weight $ ds^2_{i \sigma} > 0 $, describe the creation of excitations at the resonance frequency, while a negative weight, $ ds^2_{i \sigma} < 0 $, signifies annihilation of excitations. Note that in the response negative dips at positive frequencies were recently connected with excited states of the system~\cite{Scarlatella_2019}. 

We would like to emphasize the following: the sign of the norm $ds^2_{i \sigma}$ encapsulates an important physical meaning, i.e., it determines whether the dominant processes at the resonance frequency are particle- or holelike. These two types of processes can be regarded as the light-matter analogies of particles and holes in condensed-matter systems. Stressing this analogy further, the peak swap between the two NPs above threshold signals a transition from particlelike to bosonic-hole-like physics at positive frequencies and vice versa for negative frequencies. The e-NP can now be understood as a population-inverted state~\cite{Scarlatella_2019} whose response in the closed system favors the creation of cavity excitations to lower its energy and eventually decay to the SP~\cite{Scully}. Consequently, the NP$\to$e-NP transition does not survive in the closed system.

\section{Open system dynamics}
Infinitesimal dissipation stabilizes the e-NP, thereby enabling the aforementioned transition. In the e-NP, the system absorbs photons in order to decay to the superradiant ground state [see Fig.~\ref{fig:open_final}(a), red arrows] and dissipation acts as an incoherent drive realizing a population-inverted state into an empty cavity [see Fig.~\ref{fig:open_final}(a), orange arrows]. To further understand the nature of this transition, we use \textit{third-quantization} to obtain the open-system counterparts of $\omega_1, \omega_2$, namely, the Liouvillian eigenvalues, $\mathcal{\epsilon}_{1,2}$ - complex numbers whose imaginary (real) parts describes the frequencies (lifetimes) of the excitations~\cite{Prosen_2010}. Convergence to the steady state necessitates $ \re({\epsilon_{1,2}}) <0$, as the Liouvillian eigensolutions evolve according to $e^{\epsilon_{1,2} t}$.

Similar to Fig.~\ref{fig:closed_sys}(b), the high-energy mode ${\mathcal{\epsilon}}_1$ (corresponding to $\pm \omega_1$ in the closed system) is represented by a two complex-conjugated eigenvalues pair that does not show any significant feature (not plotted). In stark contrast, the soft mode pair, $\mathcal{\epsilon}_2$ (corresponding to $\pm\omega_2$ in the closed system), exhibits much richer features, see Fig.~\ref{fig:open_final}(b). In the NP, $\mathcal{\epsilon}_2$ appears as a pair of complex-conjugated eigenvalues with degenerate negative real parts. An increase in the couplings leads to the emergence of a dissipation-stabilized exceptional point at $\lambda_3^c$, where the imaginary parts coalesce to zero while the real parts split. This exceptional point is responsible for the soft peak merger in the dynamical response, see Fig.~\ref{fig:open_final}(c). The situation evolves specularly until $\lambda_4^c$, where the system reaches the stable e-NP phase. $\lambda_3^c$ and $\lambda_4^c$ coincide with the locations where instabilities to the SP occur in the closed system, cf.~Fig.~\ref{fig:closed_sys}(b). Dissipation lifts this instability by preventing the real part of the eigenvalues from becoming positive.

We cannot associate a norm [cf.~Eq.~\eqref{eq:symplectic_norm}] to the open-system modes. Luckily, as we show above, the norm dictates the signs in the dynamical response [cf.~Eq.~\eqref{eq:NMresponse}]. Hence, we can fully characterize the particle- or holelike nature of the excitations by studying the signs and full-widths at half-maxima (FWHM) of the peaks (positive peaks correspond to particle-like frequencies). The real part of the Liouvillian eigenvalues dictates the width of the peaks in $\mathcal{A}(\omega)$, showing that one peak is overdamped with respect to the other, see Fig.~\ref{fig:open_final}(b) bottom panel. Focusing on the positive peak (particlelike) associated with the radial mode [Fig.~\ref{fig:open_final}(c)], we see that its FWHM first decreases towards the exceptional point $\lambda_3^c$, signaling that the peak becomes sharper, in accord with an enhanced life-time of the state. A similar behavior is observed as one approaches the exceptional point at $\lambda_4^c$ from the stable e-NP phase. This allows us to unambiguously assign the particle- or holelike labels to the open system eigenfrequencies even beyond, $\lambda_3^c$, see Figs.~\ref{fig:open_final}(b)-\ref{fig:open_final}(d).

Our results are summarized in the new dynamical phase diagram shown in Fig.~\ref{fig:open_final}(d). Region I is characterized by the standard NP, where the soft mode's ($\epsilon_2$) particlelike excitations appears at positive frequencies, while region III is associated with the e-NP hosting particlelike excitations at negative frequencies. Performing a parameter sweep connecting the two normal phases through the new region II, a scenario reminiscent of exceptional points is realized. Though static observables remain unchanged across this sweep, the dynamical response function is strongly modified. The response exhibits a merger of particle- and holelike peaks at the first boundary ($\lambda_3^c$) and an inversion and splitting of these peaks at the second boundary ($\lambda_4^c$). In the intermediate region II, where the two peaks are merged, the particlelike response is squeezed (becomes narrower) while the holelike response broadens.

\section{Conclusions}
We find that dissipation introduces a scenario where two phases that exhibit identical stationary features differ by a dynamical quantity, thus unveiling a particle-to-hole inversion in a bosonic system. Dissipation acts as an incoherent drive that induces a population inversion into a dark cavity. This scenario is akin to hole lasing, which we believe can have important implications. Importantly, in the presence of stochastic noise sources, we postulate that the physics discussed in our work provides a putative explanation of the dissipation-generated slow Markovian cascade between SP phases seen in recently reported experiments of a BEC in cavities~\cite{Dogra_2019,Zhiqiang_2017}.

\begin{acknowledgments}
We thank S. Diehl and I. Carusotto for fruitful discussions. We acknowledge financial support from the Swiss National Science Foundation (PP00P2\_163818).
\end{acknowledgments}

\appendix
\section{Holstein-Primakoff's representation}
\label{appendix_HP}

In this Appendix, we provide the details of the Holstein-Primakoff representation for the spin operators. The Holstein-Primakoff representation for the spins is $S_+= b^\dagger \sqrt{N - b^\dagger b}$ and $S_z= -\frac{N}{2} + b^\dagger b$. We start from the Hamiltonian Eq.~\eqref{eq:IDTC},
\begin{align}
 \label{eq_sm:IDTC}
 H = & \hbar\omega_c a^\dagger a + \hbar\omega_a S_z \nonumber \\
 		& + {\frac{2\hbar\lambda_x}{\sqrt{N}}} S_x (a + a^\dagger)+ {\frac{2\hbar\lambda_y}{\sqrt{N}}} i S_y (a - a^\dagger)\,,
\end{align}
and expand around the respective mean-field solutions, $ a = \alpha\sqrt{2N} + c$ and $b = \beta\sqrt{2N} + d$, where $\alpha$ and $\beta$ are complex numbers defining coherent states. This leads to a fluctuation Hamiltonian describing two coupled bosonic modes of the form Eq.~\eqref{eq:fluc_hamiltonian}, with the coefficients defined as~\cite{Soriente_2018}
\small
\begin{align}
	  \delta\bar{\omega}_1 & =-\frac{4}{\sqrt{1 - |\beta|^2}}\Bigg(1 + \frac{|\beta|^2}{4 \big(1 - |\beta|^2\big)}\Bigg) \nonumber \\ 
	  & \quad\,\,\times\Bigg[ \betare \alphare \lambda_x + \betaim \alphaim\lambda_y \Bigg]\,, \label{eq_sm:do1} \\
	  \delta\bar{\omega}_2 & = - \frac{2 \beta^*}{\sqrt{1 - |\beta|^2}}\Bigg[ \alphare\bigg(1 + \frac{\betare}{2(1 - |\beta|^2)}\beta^*\bigg) \lambda_x + \nonumber \\
	  & \quad\,\,- i \alphaim\bigg(1 + i \frac{\betaim}{2(1 - |\beta|^2)}\beta^*\bigg) \lambda_y \Bigg]\,, \label{eq_sm:do2} \\
	  \bar{\lambda}_1 & = (\lambda_x + \lambda_y) \sqrt{ 1 - |\beta|^2} -\frac{\beta}{\sqrt{1 - |\beta|^2}}\big(\lambda_x \betare - i \lambda_y \betaim\big)\,, \label{eq_sm:dl1} \\
	  \bar{\lambda}_2 & = (\lambda_x - \lambda_y) \sqrt{ 1 - |\beta|^2} -\frac{\beta}{\sqrt{1 - |\beta|^2}}\big(\lambda_x \betare + i \lambda_y \betaim\big) \,.\label{eq_sm:dl2}
\end{align}
\normalsize

%%%%%%%%%%%%%%%%%%%%%%%%%%%%%%%%%%%%%%%%%%%%%%%%%%%%%%%%%%%%%%%%%%%%%%%%%%%%%%%%%%%%%%%%%%%%%%%%
%%%%%%%%%%%%%%%%%%%%%%%%%%%%%%%%%%%%%%%%%%%%%%%%%%%%%%%%%%%%%%%%%%%%%%%%%%%%%%%%%%%%%%%%%%%%%%%%
%%%%%%%%%%%%%%%%%%%%%%%%%%%%%%%%%%%%%%%%%%%%%%%%%%%%%%%%%%%%%%%%%%%%%%%%%%%%%%%%%%%%%%%%%%%%%%%%

\section{Mean-field solution}
\label{appendix_MF}

We here provide the mean-field equations describing the IDTC model, Eq.~\eqref{eq:IDTC}. The master equation governing the evolution of the density matrix of the system is
\begin{equation}
    \label{eq_sm:Liouvillian}
    \frac{d \rho_{\rm sys}}{dt} = - \frac{i}{\hbar} [H(t), \rho_{\rm sys}]  + \kappa[ 2 a \rho_{\rm sys} a^\dagger - \{ a^\dagger a , \rho_{\rm sys}\}]\,,
\end{equation}
with $H$ the Hamiltonian Eq.~\eqref{eq:IDTC} and $\kappa$ the cavity decay rate. The equation of motion governing the mean-field order parameters are
\begin{gather}
\omega_c\alphaim - \kappa \alphare -2\lambda_y Y = 0\,,  \\ %\label{eq_sm:MF1}
\omega_c \alphare +\kappa\alphaim +  2 \lambda_x X  = 0\,, \\ %\label{eq_sm:MF2}
 \omega_a Y + 4 \lambda_y \alphaim  Z  = 0\,,  \\ %\label{eq_sm:MF3}
 \omega_a X - 4 \lambda_x \alphare Z  = 0\,,  %\label{eq_sm:MF4}
\end{gather}
where we defined $\langle a \rangle = \sqrt{N} \alpha$, $\langle S_x\rangle = NX$, $\langle S_y\rangle = NY$, and $\langle S_z\rangle = NZ$ and have taken the steady-state limit. We are interested in the description of the normal phase that coincide with the trivial solution of this system, i.e., $\alphare = \alphaim = 0$, $X = Y = 0$, and $Z = -1/2$. The superradiant phase can be also analytical obtained~\cite{Soriente_2018}. The closed-system evolution can be recovered by taking $\kappa \rightarrow 0$.

%%%%%%%%%%%%%%%%%%%%%%%%%%%%%%%%%%%%%%%%%%%%%%%%%%%%%%%%%%%%%%%%%%%%%%%%%%%%%%%%%%%%%%%%%%%%%%%%
%%%%%%%%%%%%%%%%%%%%%%%%%%%%%%%%%%%%%%%%%%%%%%%%%%%%%%%%%%%%%%%%%%%%%%%%%%%%%%%%%%%%%%%%%%%%%%%%
%%%%%%%%%%%%%%%%%%%%%%%%%%%%%%%%%%%%%%%%%%%%%%%%%%%%%%%%%%%%%%%%%%%%%%%%%%%%%%%%%%%%%%%%%%%%%%%%

\section{Keldysh action}
\label{appendix_KA}

In this Appendix, we provide details on the Keldysh treatment of the open IDTC model and how one can calculate the Green's functions of the system. We introduce the combined Nambu--Keldysh spinor in the rotated classical and quantum basis,
\begin{align}
\label{TCNambuKeldyshspinorSR}
    \delta\Phi(\omega) & = (\delta a_c(\omega) \delta a_c^*(-\omega) \delta b_c(\omega) \delta b_c^*(-\omega) \delta a_q(\omega) \nonumber \\
    & \quad\, \times \delta a_q^*(-\omega) \delta b_q(\omega) \delta b_q^*(-\omega))^\intercal, \nonumber\\
    \delta\Phi^\dagger(\omega) & = (\delta a_c^*(\omega) \delta a_c(-\omega) \delta b_c^*(\omega) \delta b_c(-\omega) \delta a_q^*(\omega) \nonumber \\
    & \quad \, \times \delta a_q(-\omega) \delta b_q^*(\omega) \delta b_q(-\omega)),
\end{align}
and write the quadratic action expressed on the Keldysh contour in frequency domain as
\begin{equation}
\label{TCSRNambuKeldyshaction}
  S = \frac{1}{2}\int_\omega \; \delta\Phi^\dagger(\omega)\begin{pmatrix} 0 & {\big[G_{4\times 4}^A\big]}^{-1} \\
                          {\big[G_{4\times 4}^R\big]}^{-1} & {D_{4\times 4}^K} \end{pmatrix} \delta\Phi(\omega),
\end{equation}
with the inverse Green's functions
\begin{widetext}
\begin{align}
  {\big[G_{4\times 4}^R\big]}^{-1} & = \begin{pmatrix} \omega -\omega_c + i\kappa &         0         &        -\bar{\lambda}_1^*          &        -\bar{\lambda}_2             \\
                                                                     0          & -\omega -\omega_c - i\kappa &        -\bar{\lambda}_2^*          &        -\bar{\lambda}_1             \\
                                                      -\bar{\lambda}_1     &       -\bar{\lambda}_2      & \omega - \omega_a - \delta\bar{\omega}_1   &    -\delta\bar{\omega}_2^*          \\
                                                      -\bar{\lambda}_2^*   &       -\bar{\lambda}_1^*      &     -\delta\bar{\omega}_2        & -\omega - \omega_a - \delta\bar{\omega}_1^*
                                                                                \end{pmatrix}\,, \\
  {\big[G_{4\times 4}^A\big]}^{-1} & = \begin{pmatrix} \omega -\omega_c - i\kappa &         0         &        -\bar{\lambda}_1^*          &        -\bar{\lambda}_2             \\
                                                                     0          & -\omega -\omega_c + i\kappa &        -\bar{\lambda}_2^*          &        -\bar{\lambda}_1             \\
                                                      -\bar{\lambda}_1     &       -\bar{\lambda}_2      & \omega - \omega_a - \delta\bar{\omega}_1^*   &    -\delta\bar{\omega}_2^*          \\
                                                      -\bar{\lambda}_2^*   &       -\bar{\lambda}_1^*      &     -\delta\bar{\omega}_2        & -\omega - \omega_a - \delta\bar{\omega}_1
                                                                  \end{pmatrix}\,,
\end{align}
\end{widetext}
and the Keldysh self-energy
\begin{equation}
  D_{4\times 4}^K = 2i\text{ diag}(\kappa,\kappa,0,0).
\end{equation}
The action $S$ is quadratic in both the cavity and the bosonized spin fields. This allows us to integrate out the spin degrees of freedom (the $d$ operators) and obtain a photon only action. We do this according to standard Gaussian integration
\begin{align}
	\int & \mathcal{D}\left[\psi,\psi^\dagger\right] \; e^{i\int_q \psi^\dagger\left(q\right)D\left(q\right)\psi\left(q\right) + i\int_q \left(\phi^\dagger\left(q\right)\psi\left(q\right) + \psi^\dagger\left(q\right)\chi\left(q\right)\right)} \nonumber \\
	& = \left(\det D\right)^{-1} e^{i\int_q \phi^\dagger\left(q\right)D^{-1}\left(q\right)\chi\left(q\right)}\,,
\end{align}
to obtain
\begin{widetext}
\begin{equation}
	S_{photon}[\delta a^*,\delta a] = \int_\omega \delta A_4^\dagger(\omega)\begin{pmatrix} 0 & \big[G_{2\times 2}^{A,p}\big]^{-1}(\omega)  \\
                                    \big[G_{2\times 2}^{R,p}\big]^{-1}(\omega) & D_{2\times 2}^{K,p}(\omega) \end{pmatrix} \delta A_4(\omega). \label{photononlyactionSRTC}
\end{equation}
The various terms correspond to the photon four-vector that collects the classical and quantum field components
\begin{equation}
  \delta A_4(\omega) = \begin{pmatrix}  \delta a_c(\omega)  \\  \delta a_c^*(-\omega) \\  \delta a_q(\omega)  \\  \delta a_q^*(-\omega) \end{pmatrix},
\end{equation}
and the block entries are $2\times2$ inverse photon Green's functions.
\begin{equation}
  \label{eq_sm:photon_GR}
  {\big[G_{2\times 2}^{R,p}\big]}^{-1} =  \begin{pmatrix} \omega -\omega_c + i\kappa  + \Sigma_1^{R,p}(\omega)  & \Sigma_{2}^{R,p}(\omega)  \\
                                          {\Sigma_{2}^{R,p}}^*(-\omega) & -\omega -\omega_c - i\kappa + {\Sigma_{1}^{R,p}}^*(-\omega) \end{pmatrix}
\end{equation}
\end{widetext}
where the $\Sigma_i$ are self-energies originating from the integration. As always ${[G_{2\times 2}^{A,p}]}^{-1} = {({[G_{2\times 2}^{R,p}]}^{-1})}^{\dagger}$. The Keldysh component of the photon action is
\begin{equation}
  D^K = 2i\kappa \mathcal{I}.
\end{equation}

%%%%%%%%%%%%%%%%%%%%%%%%%%%%%%%%%%%%%%%%%%%%%%%%%%%%%%%%%%%%%%%%%%%%%%%%%%%%%%%%%%%%%%%%%%%%%%%%
%%%%%%%%%%%%%%%%%%%%%%%%%%%%%%%%%%%%%%%%%%%%%%%%%%%%%%%%%%%%%%%%%%%%%%%%%%%%%%%%%%%%%%%%%%%%%%%%
%%%%%%%%%%%%%%%%%%%%%%%%%%%%%%%%%%%%%%%%%%%%%%%%%%%%%%%%%%%%%%%%%%%%%%%%%%%%%%%%%%%%%%%%%%%%%%%%

\section{Superradiant Phase}
\label{appendix_SR}

\begin{figure}[h!]
  \includegraphics[width=\columnwidth]{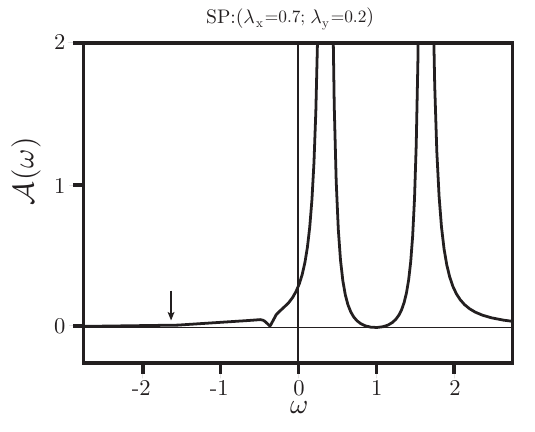}
  \caption{Cavity dynamical response function over the SP, $\lambda_{SP}=(0.7,0.2)$. The response presents four peaks appearing at paired frequencies (arrow indicates unresolved small peak). Positive (negative) frequencies correspond to positive (negative) peaks, as in the NP [cf. Fig.~1(b) main body], $\omega_c=\omega_a,\kappa/\omega_c= 0.1$}.
  \label{fig:supmat}
\end{figure}

In this Appendix, we present results regarding the response function in the superradiant phase. The cavity response of the superradiant phase can be readily calculated from Eq.~\eqref{eq_sm:photon_GR} using the coefficients Eqs.~\eqref{eq_sm:do1}-\eqref{eq_sm:dl2} with the superradiant mean-field solution derived in Ref.~\cite{Soriente_2018}. As can be seen from Fig.~\ref{fig:supmat} the SP does not present any peak inversion or extraordinary features.

%%%%%%%%%%%%%%%%%%%%%%%%%%%%%%%%%%
%%%%%%%%%%%%%%%%%%%%%%%%%%%%%%%%%

%%%%%%%%%%%%%%%%%%%%%%%%%%%%%%%%%%%%%%%%%%%%%%%%%%%%%%%%%%%%%%%%%%%%%%%%%%%%%%%%%%%%%%%%%%%%%%%%
%%%%%%%%%%%%%%%%%%%%%%%%%%%%%%%%%%%%%%%%%%%%%%%%%%%%%%%%%%%%%%%%%%%%%%%%%%%%%%%%%%%%%%%%%%%%%%%%
%%%%%%%%%%%%%%%%%%%%%%%%%%%%%%%%%%%%%%%%%%%%%%%%%%%%%%%%%%%%%%%%%%%%%%%%%%%%%%%%%%%%%%%%%%%%%%%%

\section{Normal-mode transformation}
\label{appendix_NM}

In this Appendix, we provide details on the normal-mode transformation employed to diagonalize the closed-system Hamiltonian Eq.~\eqref{eq:IDTC}. We start from the Holstein-Primakoff fluctuation Hamiltonian~\eqref{eq:fluc_hamiltonian} and recast it in the following matrix form in the operator basis $() c d c^\dagger d^\dagger )^T$:
\begin{equation}
  H_{\rm fl} = \begin{pmatrix} \omega_c  &   \bar{\lambda}_1^*   &   0   &   \bar{\lambda}_2   \\
                      \bar{\lambda}_1   & \Omega   &   \bar{\lambda}_2   &   \delta\bar{\omega}_2^* \\
                      0   &   \bar{\lambda}_1^*  &   \omega_c   &   \bar{\lambda}_1   \\
                      \bar{\lambda}_2^*   & \delta\bar{\omega}_2  &   \bar{\lambda}_1^*   &   \Omega
      \end{pmatrix}\,,
\end{equation}
where for brevity we define $\Omega = \omega_a + \delta\bar{\omega}_1$. The associated dynamical matrix to be diagonalized is~\cite{Xiao_2009}
\begin{equation}
  D = \begin{pmatrix} \omega_c  &   \bar{\lambda}_1^*   &   0   &   \bar{\lambda}_2   \\
                      \bar{\lambda}_1   & \Omega   &   \bar{\lambda}_2   &   \delta\bar{\omega}_2^* \\
                      0   &   -\bar{\lambda}_1^*  &   -\omega_c   &   -\bar{\lambda}_1   \\
                      -\bar{\lambda}_2^*   & -\delta\bar{\omega}_2  &   -\bar{\lambda}_1^*   &   -\Omega
      \end{pmatrix}\,.
\end{equation}
We need to ensure that after diagonalization the newly found operators, i.e., the normal modes, are true bosonic operators, namely, that they satisfy bosonic commutation relations. To achieve that, we need to impose some constraints on the actual form of the eigenvectors. This is done by imposing constraints on their normalization and on the structure of the transformation matrix, $T$, that diagonalizes $D$.
Solving the eigenvalue problem
\begin{equation}
  D \mathcal{T} = \omega \mathcal{T}\,,
\end{equation}
for the eigenvector $\mathcal{T}$ we find that the eigenvalues of $D$ are
\begin{equation}
  \omega_{\pm}^2 = \frac{-m \pm \sqrt{m^2 - 4n}}{2}\,,
\end{equation}
with
\begin{widetext}
\begin{gather}
  m = - \omega_c^2 - \Omega^2 - 2 \left|\bar{\lambda}_1\right|^2 + 2 \left|\bar{\lambda}_2\right|^2 + \left|\delta\bar{\omega}_2\right|^2\,, \\
  n = \left( \left|\bar{\lambda}_1\right|^2 - \left|\bar{\lambda}_2\right|^2 \right)^2 + \omega_c^2 \left( \Omega^2 - \left|\delta\bar{\omega}_2\right|^2 \right) + 4 \omega_c \re{\left[ \bar{\lambda}_1 \bar{\lambda}_2 \delta\bar{\omega}_2^* \right]} - 2 \omega_c \Omega \left( \left|\bar{\lambda}_1\right|^2 + \left|\bar{\lambda}_2\right|^2 \right)\,. 
\end{gather}
We define the bare eigenfrequencies as the positive ones,
\begin{align}
  & \omega_1 = +\left|\omega_+\right|\,,  \\
  & \omega_2 = +\left|\omega_-\right|\,. 
\end{align}
Before writing the Bogoliubov transformation matrix that maps to the normal modes, we introduce some useful functions:
\begin{gather}
  t_c^+(\omega) =  \frac{\bar{\lambda}_1^*}{\omega - \omega_c} t_d^+(\omega) + \frac{\bar{\lambda}_2}{\omega - \omega_c}\,, \\
  t_d^+(\omega) =  - \frac{2\omega_c \bar{\lambda}_1 \bar{\lambda}_2 + \delta\bar{\omega}_2^* \left(\omega^2 - \omega_c^2\right)}{\left|\bar{\lambda}_1\right|^2 \left(\omega + \omega_c\right) - \left|\bar{\lambda}_2\right|^2 \left(\omega - \omega_c\right) + \left(\Omega - \omega\right) \left(\omega^2 - \omega_c^2\right)}\,, \\
  t_c^-(\omega) =  - \frac{\bar{\lambda}_2^*}{\omega + \omega_c} t_d^+(\omega) - \frac{\bar{\lambda}_1}{\omega + \omega_c}\,, \\
  t_d^-(\omega) =  1\,.
\end{gather}
Additionally, we define a key ingredient of the transformation, the \textit{symplectic norm}, according to Eq.~\eqref{eq:symplectic_norm}:
\begin{equation}
   ds^2_{i \sigma} = \left|t_c^+(\sigma{\omega}_i)\right|^2 + \left|t_d^+(\sigma{\omega}_i)\right|^2 - \left|t_c^-(\sigma{\omega}_i)\right|^2 - \left|t_d^-(\sigma{\omega}_i)\right|^2 \equiv \sum_{j=\{c,d\},\sigma'}\sigma' |t_{j}^{\sigma'}(\sigma\omega_i)|^2\,,
\end{equation}
with $\sigma,\sigma' = \pm; i = 1,2$.

The transformation matrix $T$ is then found according to the following construction:
\begin{align}
  T & = \begin{bmatrix} \mathcal{T}_1 & \mathcal{T}_2 & \mathcal{T}_3 & \mathcal{T}_4 \end{bmatrix} \notag \\
  & = \begin{bmatrix} \mathcal{T}(s_{1 +}\omega_1) & \mathcal{T}(s_{2 +}\omega_2) & \mathcal{T}(s_{1 -}\omega_1) & \mathcal{T}(s_{2 -}\omega_2) \end{bmatrix} \notag \\
  & = \begin{pmatrix} \frac{t_c^+(s_{1 +}\omega_1)}{\sqrt{|ds^2_{1 +}|}} & \frac{t_c^+(s_{2 +}\omega_2)}{\sqrt{|ds^2_{2 +}|}} & \frac{t_c^+(s_{1 -}\omega_1)}{\sqrt{|ds^2_{1 -}|}} & \frac{t_c^+(s_{2 -}\omega_2)}{\sqrt{|ds^2_{2 -}|}} \\
  \frac{t_d^+(s_{1 +}\omega_1)}{\sqrt{|ds^2_{1 +}|}} & \frac{t_d^+(s_{2 +}\omega_2)}{\sqrt{|ds^2_{2 +}|}} & \frac{t_d^+(s_{1 -}\omega_1)}{\sqrt{|ds^2_{1 -}|}} & \frac{t_d^+(s_{2 -}\omega_2)}{\sqrt{|ds^2_{2 -}|}} \\
  \frac{t_c^-(s_{1 +}\omega_1)}{\sqrt{|ds^2_{1 +}|}} & \frac{t_c^-(s_{2 +}\omega_2)}{\sqrt{|ds^2_{2 +}|}} & \frac{t_c^-(s_{1 -}\omega_1)}{\sqrt{|ds^2_{1 -}|}} & \frac{t_c^-(s_{2 -}\omega_2)}{\sqrt{|ds^2_{2 -}|}} \\
  \frac{t_d^-(s_{1 +}\omega_1)}{\sqrt{|ds^2_{1 +}|}} & \frac{t_d^-(s_{2 +}\omega_2)}{\sqrt{|ds^2_{2 +}|}} & \frac{t_d^-(s_{1 -}\omega_1)}{\sqrt{|ds^2_{1 -}|}} & \frac{t_d^-(s_{2 -}\omega_2)}{\sqrt{|ds^2_{2 -}|}}
    \end{pmatrix} \,,
\end{align}
\end{widetext}
where the $\mathcal{T}$ are vectors constructed as follows:
\begin{align}
    \mathcal{T}&(s_{i \sigma}\omega_i) \nonumber \\ 
    & = \frac{1}{\sqrt{|ds^2_{i \sigma}|}}\begin{pmatrix} t_c^+(s_{i \sigma}\omega_i) & t_d^+(s_{i \sigma}\omega_i) & t_c^-(s_{i \sigma}\omega_i) & t_d^-(s_{i \sigma}\omega_i) \end{pmatrix}^\intercal\,,
\end{align}
with $s_{i \sigma} = \text{sgn}(ds^2_{i \sigma})$ and $i=1,2,\sigma=\pm$. The argument $(s_{i \sigma}\omega_i)$ ensures that the $\mathcal{T}$ components are evaluated at the eigenfrequencies with positive symplectic norm. For completeness, we highlight that the $\mathcal{T}$ vectors are pairwise related via
\begin{align}
  & \mathcal{T}(-\omega) = \Sigma_x \mathcal{T}(\omega)^*\,, \\
  & \Sigma_x = \begin{pmatrix} 0 & \mathcal{I} \\ \mathcal{I} & 0 \end{pmatrix}\,,
\end{align}
with $\mathcal{I}$ the $2\times2$ identity matrix. This allows us to simplify the notation in Eq.~\eqref{eq:NMresponse}. The $\mathcal{T}$ vectors are normalized according to
\begin{gather}
   \mathcal{T}_{1,2}^\intercal \mathcal{I}_- \mathcal{T}_{1,2} = \mathcal{T}(s_{i +}\omega_i)^\intercal \mathcal{I}_- \mathcal{T}(s_{i +}\omega_i) +1\,, \\
   \mathcal{T}_{3,4}^\intercal \mathcal{I}_- \mathcal{T}_{3,4} = \mathcal{T}(s_{i -}\omega_i)^\intercal \mathcal{I}_- \mathcal{T}(s_{i -}\omega_i) = -1 \,, \\
  \mathcal{I}_- = \begin{pmatrix} \mathcal{I} & 0 \\ 0 & -\mathcal{I} \end{pmatrix}\,,
\end{gather}
with $i=1,2$. These normalization conditions imply that the ordering of the vectors in $T$ is: such that the first two vector columns, $\mathcal{T}_{1,2}$, are always normalized to $1$ while the last two, $\mathcal{T}_{3,4}$, are normalized to $-1$, thus ensuring canonical commutation relations for the new eigenmodes. The frequencies at which the vectors element are evaluated change throughout the parameter region, $\left(\lambda_x,\lambda_y\right)$. In particular, as highlighted in the main body of the manuscript, carrying out the normal-mode transformation throughout the parameter space, we obtain that in the NP and SP phases $\mathcal{T}_{1},\mathcal{T}_{3}$ are associated with the frequencies $\omega_1, -\omega_1$, respectively, and $\mathcal{T}_{2},\mathcal{T}_{4}$ are associated with the frequencies $\omega_2, -\omega_2$. In the e-NP nothing changes for $\mathcal{T}_{1},\mathcal{T}_{3}$, but we have a frequency inversion for $\mathcal{T}_{2},\mathcal{T}_{4}$, namely, $\mathcal{T}_{2}$ is now associated with $-\omega_2$ and $\mathcal{T}_{4}$ to $\omega_2$.

With the normal mode transformation $T$
\begin{equation}
  \label{eq_sm:NMtransf}
    \phi = T \psi\,,
\end{equation}
where $\phi = \begin{pmatrix} c & d & c^\dagger & d^\dagger \end{pmatrix}^\intercal$ and $\psi = \begin{pmatrix} A_1 & A_2 & A_1^\dagger & A_2^\dagger \end{pmatrix}^\intercal$, we can express cavity and spin operators in terms of the normal modes, $A_{1,2}$, or vice versa
\begin{widetext}
\begin{align}
	& c = \frac{t_c^+(s_{1 +}\omega_1)}{\sqrt{|ds^2_{1 +}|}} A_1 + \frac{t_c^+(s_{2 +}\omega_2)}{\sqrt{|ds^2_{2 +}|}} A_2 + \frac{t_c^+(s_{1 -}\omega_1)}{\sqrt{|ds^2_{1 -}|}} A_1^\dagger + \frac{t_c^+(s_{2 -}\omega_2)}{\sqrt{|ds^2_{2 -}|}} A_2^\dagger = \frac{t_c^+(s_{1 +}\omega_1)}{\sqrt{|ds^2_{1 +}|}} A_1 + \frac{t_c^+(s_{2 +}\omega_2)}{\sqrt{|ds^2_{2 +}|}} A_2 \nonumber \\
	& \qquad  + \frac{t_c^-(s_{1 +}\omega_1)^*}{\sqrt{|ds^2_{1 +}|}} A_1^\dagger + \frac{t_c^-(s_{2 +}\omega_2)^*}{\sqrt{|ds^2_{2 +}|}} A_2^\dagger\,, \notag \\
	& c^\dagger = \frac{t_c^-(s_{1 +}\omega_1)}{\sqrt{|ds^2_{1 +}|}} A_1 + \frac{t_c^-(s_{2 +}\omega_2)}{\sqrt{|ds^2_{2 +}|}} A_2 + \frac{t_c^+(s_{1 +}\omega_1)^*}{\sqrt{|ds^2_{1 +}|}} A_1^\dagger + \frac{t_c^+(s_{2 +}\omega_2)^*}{\sqrt{|ds^2_{2 +}|}} A_2^\dagger\,,
\end{align}
where in the first line we used the relation $\mathcal{T}(-\omega) = \Sigma_x \mathcal{T}(\omega)^*$.

%%%%%%%%%%%%%%%%%%%%%%%%%%%%%%%%%%%%%%%%%%%%%%%%%%%%%%%%%%%%%%%%%%%%%%%%%%%%%%%%%%%%%%%%%%%%%%%%%%%%

\section{Response function}
\label{appendix_RF}

In this Appendix, we calculate the retarded Green's function of the closed-system Hamiltonian Eq.~\eqref{eq:fluc_hamiltonian}. For ease of notation, we define $\tilde{\omega}_i = s_{i +}\omega_i$ and with the above normal-mode transformation $T$ we have
\begin{align}
  \resizebox{.1\hsize}{!}{$\left[c\left(t\right), c^\dagger\left(t'\right)\right]$} & \resizebox{.9\hsize}{!}{$= \frac{|t_c^+(\tilde{\omega}_1)|^2}{|ds^2_{1 +}|} \left[A_1\left(t\right), A_1^\dagger\left(t'\right)\right] - \frac{|t_c^-(\tilde{\omega}_1)|^2}{|ds^2_{1 +}|} \left[A_1\left(t'\right), A_1^\dagger\left(t\right)\right] + \frac{|t_c^+(\tilde{\omega}_2)|^2}{|ds^2_{2 +}|} \left[A_2\left(t\right), A_2^\dagger\left(t'\right)\right] - \frac{|t_c^-(\tilde{\omega}_2)|^2}{|ds^2_{2 +}|} \left[A_2\left(t\right), A_2^\dagger\left(t'\right)\right]$} \notag \\
  & \resizebox{.9\hsize}{!}{$= \frac{1}{|ds^2_{1 +}|} \left( |t_c^+(\tilde{\omega}_1)|^2 e^{-i \tilde{\omega}_1 \left(t-t'\right)} - |t_c^-(\tilde{\omega}_1)|^2 e^{i \tilde{\omega}_1 \left(t-t'\right)}\right) + \frac{1}{|ds^2_{2 +}|} \left( |t_c^+(\tilde{\omega}_2)|^2 e^{-i \tilde{\omega}_2 \left(t-t'\right)} - |t_c^-(\tilde{\omega}_2)|^2 e^{i \tilde{\omega}_2 \left(t-t'\right)}\right)\,.$}
\end{align}
From the definition of the retarded Green's function $G^R(t-t')=-i\theta(t-t')\langle\left[c(t), c^\dagger(t')\right]\rangle$, the integral representation of the Heaviside function $\theta\left(\tau\right) = \lim_{\varepsilon\to 0^+} \frac{1}{2\pi i} \int_{-\infty}^{\infty} \frac{1}{x - i\varepsilon} e^{i \tau x} d x $, and by plugging in the above transformation we obtain
\begin{align}
  G^R(t,t') = & -i \lim_{\varepsilon\to 0^+} \frac{1}{2\pi i} \int_{-\infty}^{\infty} \frac{1}{x - i\varepsilon} e^{i \tau x} d x \left[ \frac{1}{|ds^2_{1 +}|} \left( |t_c^+(\tilde{\omega}_1)|^2 e^{-i\tilde{\omega}_1 \tau} - |t_c^-(\tilde{\omega}_1)|^2 e^{i\tilde{\omega}_1 \tau} \right) \right.\nonumber \\
  & \left.+ \frac{1}{|ds^2_{2 +}|} \left( |t_c^+(\tilde{\omega}_2)|^2 e^{-i\tilde{\omega}_2 \tau} - |t_c^-(\tilde{\omega}_2)|^2 e^{i\tilde{\omega}_2 \tau} \right) \right]
\end{align}
Fourier transforming with respect to $\tau = t - t'$, we obtain
\begin{equation}
  G^R\left( \omega \right) = \lim_{\varepsilon\to 0^+} \frac{1}{|ds^2_{1 +}|} \left( \frac{|t_c^+(\tilde{\omega}_1)|^2}{\omega - \tilde{\omega}_1 + i\varepsilon} - \frac{|t_c^-(\tilde{\omega}_1)|^2}{\omega + \tilde{\omega}_1 + i\varepsilon} \right) + \frac{1}{|ds^2_{2 +}|} \left( \frac{|t_c^+(\tilde{\omega}_2)|^2}{\omega - \tilde{\omega}_2 + i\varepsilon} - \frac{|t_c^-(\tilde{\omega}_2)|^2}{\omega + \tilde{\omega}_2 +i\varepsilon} \right)\,.
\end{equation}
Here, we used that the Fourier transform, $\mathcal{F}$, of a complex exponential is a delta function, i.e., $\mathcal{F}[e^{i\tau x} e^{\mp i \tilde{\omega}_i \tau}] = 2\pi\delta\left(x  + \omega \mp \tilde{\omega}_i\right)$. Using Sokhotski–Plemelj theorem~\cite{Petruccione} for the representation of the delta function we can write
\begin{equation}
  G^R\left( \omega \right) = -i\pi \frac{1}{|ds^2_{1 +}|} \left( |t_c^+(\tilde{\omega}_1)|^2 \delta\left(\omega - \tilde{\omega}_1\right) - |t_c^-(\tilde{\omega}_1)|^2 \delta\left(\omega + \tilde{\omega}_1\right) \right) - i\pi \frac{1}{|ds^2_{2 +}|} \left( |t_c^+(\tilde{\omega}_2)|^2 \delta\left(\omega - \tilde{\omega}_2\right) - |t_c^-(\tilde{\omega}_2)|^2 \delta\left(\omega + \tilde{\omega}_2\right) \right)
\end{equation}
We note here that due to the relation $\mathcal{T}(-\omega) = \Sigma_x \mathcal{T}(\omega)^*$, we have $|t_c^-(\tilde{\omega}_2)|^2 = |t_c^+(-\tilde{\omega}_2)|^2$, which was used to compactify the notation in Eq.~$(6)$ of the main text. From this derivation, we see how terms proportional to $|t_c^+(\tilde{\omega}_i)|^2$ arise from the commutator $\left[A_i\left(t\right), A_i^\dagger\left(t'\right)\right]$, i.e., from a scenario where a quasiparticle is first created and then annihilated at a later time $t>t'$. Therefore this process is associated with particlelike physics while terms proportional to $|t_c^-(\tilde{\omega}_i)|^2$ arising from the commutator $\left[A_i\left(t'\right), A_i^\dagger\left(t\right)\right]$ can be associated to holelike physics. Moreover, by direct inspection we see that the $|t_c^+(\tilde{\omega}_1)|^2$ terms are responsible for the positive peaks in the response and the $|t_c^-(\tilde{\omega}_i)|^2$ are responsible for the negative ones.
\end{widetext}
% \bibliography{biblio}

\begin{thebibliography}{52}%
\makeatletter
\providecommand \@ifxundefined [1]{%
 \@ifx{#1\undefined}
}%
\providecommand \@ifnum [1]{%
 \ifnum #1\expandafter \@firstoftwo
 \else \expandafter \@secondoftwo
 \fi
}%
\providecommand \@ifx [1]{%
 \ifx #1\expandafter \@firstoftwo
 \else \expandafter \@secondoftwo
 \fi
}%
\providecommand \natexlab [1]{#1}%
\providecommand \enquote  [1]{``#1''}%
\providecommand \bibnamefont  [1]{#1}%
\providecommand \bibfnamefont [1]{#1}%
\providecommand \citenamefont [1]{#1}%
\providecommand \href@noop [0]{\@secondoftwo}%
\providecommand \href [0]{\begingroup \@sanitize@url \@href}%
\providecommand \@href[1]{\@@startlink{#1}\@@href}%
\providecommand \@@href[1]{\endgroup#1\@@endlink}%
\providecommand \@sanitize@url [0]{\catcode `\\12\catcode `\$12\catcode
  `\&12\catcode `\#12\catcode `\^12\catcode `\_12\catcode `\%12\relax}%
\providecommand \@@startlink[1]{}%
\providecommand \@@endlink[0]{}%
\providecommand \url  [0]{\begingroup\@sanitize@url \@url }%
\providecommand \@url [1]{\endgroup\@href {#1}{\urlprefix }}%
\providecommand \urlprefix  [0]{URL }%
\providecommand \Eprint [0]{\href }%
\providecommand \doibase [0]{http://dx.doi.org/}%
\providecommand \selectlanguage [0]{\@gobble}%
\providecommand \bibinfo  [0]{\@secondoftwo}%
\providecommand \bibfield  [0]{\@secondoftwo}%
\providecommand \translation [1]{[#1]}%
\providecommand \BibitemOpen [0]{}%
\providecommand \bibitemStop [0]{}%
\providecommand \bibitemNoStop [0]{.\EOS\space}%
\providecommand \EOS [0]{\spacefactor3000\relax}%
\providecommand \BibitemShut  [1]{\csname bibitem#1\endcsname}%
\let\auto@bib@innerbib\@empty
%</preamble>
\bibitem [{\citenamefont {Breuer}\ and\ \citenamefont
  {Petruccione}(2002)}]{Petruccione}%
  \BibitemOpen
  \bibfield  {author} {\bibinfo {author} {\bibfnamefont {H.-P.}\ \bibnamefont
  {Breuer}}\ and\ \bibinfo {author} {\bibfnamefont {F.}~\bibnamefont
  {Petruccione}},\ }\href@noop {} {\emph {\bibinfo {title} {The Theory of Open
  Quantum Systems}}}\ (\bibinfo  {publisher} {Oxford University Press},\
  \bibinfo {year} {2002})\BibitemShut {NoStop}%
\bibitem [{\citenamefont {Carusotto}\ and\ \citenamefont
  {Ciuti}(2013)}]{Carusotto_2013}%
  \BibitemOpen
  \bibfield  {author} {\bibinfo {author} {\bibfnamefont {I.}~\bibnamefont
  {Carusotto}}\ and\ \bibinfo {author} {\bibfnamefont {C.}~\bibnamefont
  {Ciuti}},\ }\href {\doibase 10.1103/RevModPhys.85.299} {\bibfield  {journal}
  {\bibinfo  {journal} {Rev. Mod. Phys.}\ }\textbf {\bibinfo {volume} {85}},\
  \bibinfo {pages} {299} (\bibinfo {year} {2013})}\BibitemShut {NoStop}%
\bibitem [{\citenamefont {Ritsch}\ \emph {et~al.}(2013)\citenamefont {Ritsch},
  \citenamefont {Domokos}, \citenamefont {Brennecke},\ and\ \citenamefont
  {Esslinger}}]{Ritsch_2013}%
  \BibitemOpen
  \bibfield  {author} {\bibinfo {author} {\bibfnamefont {H.}~\bibnamefont
  {Ritsch}}, \bibinfo {author} {\bibfnamefont {P.}~\bibnamefont {Domokos}},
  \bibinfo {author} {\bibfnamefont {F.}~\bibnamefont {Brennecke}}, \ and\
  \bibinfo {author} {\bibfnamefont {T.}~\bibnamefont {Esslinger}},\ }\href
  {\doibase 10.1103/RevModPhys.85.553} {\bibfield  {journal} {\bibinfo
  {journal} {Rev. Mod. Phys.}\ }\textbf {\bibinfo {volume} {85}},\ \bibinfo
  {pages} {553} (\bibinfo {year} {2013})}\BibitemShut {NoStop}%
\bibitem [{\citenamefont {Baumann}\ \emph {et~al.}(2010)\citenamefont
  {Baumann}, \citenamefont {Guerlin}, \citenamefont {Brennecke},\ and\
  \citenamefont {Esslinger}}]{Baumann_2010}%
  \BibitemOpen
  \bibfield  {author} {\bibinfo {author} {\bibfnamefont {K.}~\bibnamefont
  {Baumann}}, \bibinfo {author} {\bibfnamefont {C.}~\bibnamefont {Guerlin}},
  \bibinfo {author} {\bibfnamefont {F.}~\bibnamefont {Brennecke}}, \ and\
  \bibinfo {author} {\bibfnamefont {T.}~\bibnamefont {Esslinger}},\ }\href
  {http://dx.doi.org/10.1038/nature09009} {\bibfield  {journal} {\bibinfo
  {journal} {Nature}\ }\textbf {\bibinfo {volume} {464}},\ \bibinfo {pages}
  {1301} (\bibinfo {year} {2010})}\BibitemShut {NoStop}%
\bibitem [{\citenamefont {Klinder}\ \emph {et~al.}(2015)\citenamefont
  {Klinder}, \citenamefont {Ke{\ss}ler}, \citenamefont {Wolke}, \citenamefont
  {Mathey},\ and\ \citenamefont {Hemmerich}}]{Klinder_2015}%
  \BibitemOpen
  \bibfield  {author} {\bibinfo {author} {\bibfnamefont {J.}~\bibnamefont
  {Klinder}}, \bibinfo {author} {\bibfnamefont {H.}~\bibnamefont {Ke{\ss}ler}},
  \bibinfo {author} {\bibfnamefont {M.}~\bibnamefont {Wolke}}, \bibinfo
  {author} {\bibfnamefont {L.}~\bibnamefont {Mathey}}, \ and\ \bibinfo {author}
  {\bibfnamefont {A.}~\bibnamefont {Hemmerich}},\ }\href {\doibase
  10.1073/pnas.1417132112} {\bibfield  {journal} {\bibinfo  {journal}
  {Proceedings of the National Academy of Sciences}\ }\textbf {\bibinfo
  {volume} {112}},\ \bibinfo {pages} {3290} (\bibinfo {year}
  {2015})}\BibitemShut {NoStop}%
\bibitem [{\citenamefont {Zhiqiang}\ \emph {et~al.}(2017)\citenamefont
  {Zhiqiang}, \citenamefont {Lee}, \citenamefont {Kumar}, \citenamefont
  {Arnold}, \citenamefont {Masson}, \citenamefont {Parkins},\ and\
  \citenamefont {Barrett}}]{Zhiqiang_2017}%
  \BibitemOpen
  \bibfield  {author} {\bibinfo {author} {\bibfnamefont {Z.}~\bibnamefont
  {Zhiqiang}}, \bibinfo {author} {\bibfnamefont {C.~H.}\ \bibnamefont {Lee}},
  \bibinfo {author} {\bibfnamefont {R.}~\bibnamefont {Kumar}}, \bibinfo
  {author} {\bibfnamefont {K.~J.}\ \bibnamefont {Arnold}}, \bibinfo {author}
  {\bibfnamefont {S.~J.}\ \bibnamefont {Masson}}, \bibinfo {author}
  {\bibfnamefont {A.~S.}\ \bibnamefont {Parkins}}, \ and\ \bibinfo {author}
  {\bibfnamefont {M.~D.}\ \bibnamefont {Barrett}},\ }\href {\doibase
  10.1364/OPTICA.4.000424} {\bibfield  {journal} {\bibinfo  {journal} {Optica}\
  }\textbf {\bibinfo {volume} {4}},\ \bibinfo {pages} {424} (\bibinfo {year}
  {2017})}\BibitemShut {NoStop}%
\bibitem [{\citenamefont {Morales}\ \emph {et~al.}(2019)\citenamefont
  {Morales}, \citenamefont {Dreon}, \citenamefont {Li}, \citenamefont
  {Baumg\"artner}, \citenamefont {Zupancic}, \citenamefont {Donner},\ and\
  \citenamefont {Esslinger}}]{Morales_2019}%
  \BibitemOpen
  \bibfield  {author} {\bibinfo {author} {\bibfnamefont {A.}~\bibnamefont
  {Morales}}, \bibinfo {author} {\bibfnamefont {D.}~\bibnamefont {Dreon}},
  \bibinfo {author} {\bibfnamefont {X.}~\bibnamefont {Li}}, \bibinfo {author}
  {\bibfnamefont {A.}~\bibnamefont {Baumg\"artner}}, \bibinfo {author}
  {\bibfnamefont {P.}~\bibnamefont {Zupancic}}, \bibinfo {author}
  {\bibfnamefont {T.}~\bibnamefont {Donner}}, \ and\ \bibinfo {author}
  {\bibfnamefont {T.}~\bibnamefont {Esslinger}},\ }\href {\doibase
  10.1103/PhysRevA.100.013816} {\bibfield  {journal} {\bibinfo  {journal}
  {Phys. Rev. A}\ }\textbf {\bibinfo {volume} {100}},\ \bibinfo {pages}
  {013816} (\bibinfo {year} {2019})}\BibitemShut {NoStop}%
\bibitem [{\citenamefont {Rodr\'{\i}guez~Chiacchio}\ and\ \citenamefont
  {Nunnenkamp}(2018)}]{Chiacchio_2018_2}%
  \BibitemOpen
  \bibfield  {author} {\bibinfo {author} {\bibfnamefont {E.~I.}\ \bibnamefont
  {Rodr\'{\i}guez~Chiacchio}}\ and\ \bibinfo {author} {\bibfnamefont
  {A.}~\bibnamefont {Nunnenkamp}},\ }\href {\doibase
  10.1103/PhysRevA.98.023617} {\bibfield  {journal} {\bibinfo  {journal} {Phys.
  Rev. A}\ }\textbf {\bibinfo {volume} {98}},\ \bibinfo {pages} {023617}
  (\bibinfo {year} {2018})}\BibitemShut {NoStop}%
\bibitem [{\citenamefont {Reiter}\ \emph {et~al.}(2018)\citenamefont {Reiter},
  \citenamefont {Nguyen}, \citenamefont {Home},\ and\ \citenamefont
  {Yelin}}]{Reiter_2018}%
  \BibitemOpen
  \bibfield  {author} {\bibinfo {author} {\bibfnamefont {F.}~\bibnamefont
  {Reiter}}, \bibinfo {author} {\bibfnamefont {T.~L.}\ \bibnamefont {Nguyen}},
  \bibinfo {author} {\bibfnamefont {J.~P.}\ \bibnamefont {Home}}, \ and\
  \bibinfo {author} {\bibfnamefont {S.~F.}\ \bibnamefont {Yelin}},\ }\href@noop
  {} {\bibfield  {journal} {\bibinfo  {journal} {arXiv:1807.06026}\ } (\bibinfo
  {year} {2018})}\BibitemShut {NoStop}%
\bibitem [{\citenamefont {Lambert}\ \emph {et~al.}(2019)\citenamefont
  {Lambert}, \citenamefont {Ahmed}, \citenamefont {Cirio},\ and\ \citenamefont
  {Nori}}]{Lambert_2019}%
  \BibitemOpen
  \bibfield  {author} {\bibinfo {author} {\bibfnamefont {N.}~\bibnamefont
  {Lambert}}, \bibinfo {author} {\bibfnamefont {S.}~\bibnamefont {Ahmed}},
  \bibinfo {author} {\bibfnamefont {M.}~\bibnamefont {Cirio}}, \ and\ \bibinfo
  {author} {\bibfnamefont {F.}~\bibnamefont {Nori}},\ }\href {\doibase
  10.1038/s41467-019-11656-1} {\bibfield  {journal} {\bibinfo  {journal}
  {Nature Communications}\ }\textbf {\bibinfo {volume} {10}},\ \bibinfo {pages}
  {3721} (\bibinfo {year} {2019})}\BibitemShut {NoStop}%
\bibitem [{\citenamefont {Kasprzak}\ \emph {et~al.}(2006)\citenamefont
  {Kasprzak}, \citenamefont {Richard}, \citenamefont {Kundermann},
  \citenamefont {Baas}, \citenamefont {Jeambrun}, \citenamefont {Keeling},
  \citenamefont {Marchetti}, \citenamefont {Szyma{\'n}ska}, \citenamefont
  {Andr{\'e}}, \citenamefont {Staehli},\ and\ \citenamefont
  {et~al.}}]{Kasprzak_2006}%
  \BibitemOpen
  \bibfield  {author} {\bibinfo {author} {\bibfnamefont {J.}~\bibnamefont
  {Kasprzak}}, \bibinfo {author} {\bibfnamefont {M.}~\bibnamefont {Richard}},
  \bibinfo {author} {\bibfnamefont {S.}~\bibnamefont {Kundermann}}, \bibinfo
  {author} {\bibfnamefont {A.}~\bibnamefont {Baas}}, \bibinfo {author}
  {\bibfnamefont {P.}~\bibnamefont {Jeambrun}}, \bibinfo {author}
  {\bibfnamefont {J.~M.~J.}\ \bibnamefont {Keeling}}, \bibinfo {author}
  {\bibfnamefont {F.~M.}\ \bibnamefont {Marchetti}}, \bibinfo {author}
  {\bibfnamefont {M.~H.}\ \bibnamefont {Szyma{\'n}ska}}, \bibinfo {author}
  {\bibfnamefont {R.}~\bibnamefont {Andr{\'e}}}, \bibinfo {author}
  {\bibfnamefont {J.~L.}\ \bibnamefont {Staehli}}, \ and\ \bibinfo {author}
  {\bibnamefont {et~al.}},\ }\href {\doibase 10.1038/nature05131} {\bibfield
  {journal} {\bibinfo  {journal} {Nature}\ }\textbf {\bibinfo {volume} {443}},\
  \bibinfo {pages} {409} (\bibinfo {year} {2006})}\BibitemShut {NoStop}%
\bibitem [{\citenamefont {Bohnet}\ \emph {et~al.}(2012)\citenamefont {Bohnet},
  \citenamefont {Chen}, \citenamefont {Weiner}, \citenamefont {Meiser},
  \citenamefont {Holland},\ and\ \citenamefont {Thompson}}]{Bohnet_2012}%
  \BibitemOpen
  \bibfield  {author} {\bibinfo {author} {\bibfnamefont {J.~G.}\ \bibnamefont
  {Bohnet}}, \bibinfo {author} {\bibfnamefont {Z.}~\bibnamefont {Chen}},
  \bibinfo {author} {\bibfnamefont {J.~M.}\ \bibnamefont {Weiner}}, \bibinfo
  {author} {\bibfnamefont {D.}~\bibnamefont {Meiser}}, \bibinfo {author}
  {\bibfnamefont {M.~J.}\ \bibnamefont {Holland}}, \ and\ \bibinfo {author}
  {\bibfnamefont {J.~K.}\ \bibnamefont {Thompson}},\ }\href
  {http://dx.doi.org/10.1038/nature10920} {\bibfield  {journal} {\bibinfo
  {journal} {Nature}\ }\textbf {\bibinfo {volume} {484}},\ \bibinfo {pages}
  {78} (\bibinfo {year} {2012})}\BibitemShut {NoStop}%
\bibitem [{\citenamefont {Kirton}\ and\ \citenamefont
  {Keeling}(2018)}]{Kirton_2018}%
  \BibitemOpen
  \bibfield  {author} {\bibinfo {author} {\bibfnamefont {P.}~\bibnamefont
  {Kirton}}\ and\ \bibinfo {author} {\bibfnamefont {J.}~\bibnamefont
  {Keeling}},\ }\href@noop {} {\bibfield  {journal} {\bibinfo  {journal} {New
  Journal of Physics}\ }\textbf {\bibinfo {volume} {20}},\ \bibinfo {pages}
  {015009} (\bibinfo {year} {2018})}\BibitemShut {NoStop}%
\bibitem [{\citenamefont {Laske}\ \emph {et~al.}(2019)\citenamefont {Laske},
  \citenamefont {Winter},\ and\ \citenamefont {Hemmerich}}]{Laske_2019}%
  \BibitemOpen
  \bibfield  {author} {\bibinfo {author} {\bibfnamefont {T.}~\bibnamefont
  {Laske}}, \bibinfo {author} {\bibfnamefont {H.}~\bibnamefont {Winter}}, \
  and\ \bibinfo {author} {\bibfnamefont {A.}~\bibnamefont {Hemmerich}},\
  }\href@noop {} {\bibfield  {journal} {\bibinfo  {journal} {arXiv:1903.10196}\
  } (\bibinfo {year} {2019})}\BibitemShut {NoStop}%
\bibitem [{\citenamefont {L{\'e}onard}\ \emph {et~al.}(2017)\citenamefont
  {L{\'e}onard}, \citenamefont {Morales}, \citenamefont {Zupancic},
  \citenamefont {Esslinger},\ and\ \citenamefont {Donner}}]{Leonard_2017_n}%
  \BibitemOpen
  \bibfield  {author} {\bibinfo {author} {\bibfnamefont {J.}~\bibnamefont
  {L{\'e}onard}}, \bibinfo {author} {\bibfnamefont {A.}~\bibnamefont
  {Morales}}, \bibinfo {author} {\bibfnamefont {P.}~\bibnamefont {Zupancic}},
  \bibinfo {author} {\bibfnamefont {T.}~\bibnamefont {Esslinger}}, \ and\
  \bibinfo {author} {\bibfnamefont {T.}~\bibnamefont {Donner}},\ }\href
  {http://dx.doi.org/10.1038/nature21067} {\bibfield  {journal} {\bibinfo
  {journal} {Nature}\ }\textbf {\bibinfo {volume} {543}},\ \bibinfo {pages}
  {87} (\bibinfo {year} {2017})}\BibitemShut {NoStop}%
\bibitem [{\citenamefont {Mivehvar}\ \emph {et~al.}(2018)\citenamefont
  {Mivehvar}, \citenamefont {Ostermann}, \citenamefont {Piazza},\ and\
  \citenamefont {Ritsch}}]{Mivehvar_2018}%
  \BibitemOpen
  \bibfield  {author} {\bibinfo {author} {\bibfnamefont {F.}~\bibnamefont
  {Mivehvar}}, \bibinfo {author} {\bibfnamefont {S.}~\bibnamefont {Ostermann}},
  \bibinfo {author} {\bibfnamefont {F.}~\bibnamefont {Piazza}}, \ and\ \bibinfo
  {author} {\bibfnamefont {H.}~\bibnamefont {Ritsch}},\ }\href {\doibase
  10.1103/PhysRevLett.120.123601} {\bibfield  {journal} {\bibinfo  {journal}
  {Physical Review Letters}\ }\textbf {\bibinfo {volume} {120}},\ \bibinfo
  {pages} {123601} (\bibinfo {year} {2018})}\BibitemShut {NoStop}%
\bibitem [{\citenamefont {Soriente}\ \emph {et~al.}(2018)\citenamefont
  {Soriente}, \citenamefont {Donner}, \citenamefont {Chitra},\ and\
  \citenamefont {Zilberberg}}]{Soriente_2018}%
  \BibitemOpen
  \bibfield  {author} {\bibinfo {author} {\bibfnamefont {M.}~\bibnamefont
  {Soriente}}, \bibinfo {author} {\bibfnamefont {T.}~\bibnamefont {Donner}},
  \bibinfo {author} {\bibfnamefont {R.}~\bibnamefont {Chitra}}, \ and\ \bibinfo
  {author} {\bibfnamefont {O.}~\bibnamefont {Zilberberg}},\ }\href {\doibase
  10.1103/PhysRevLett.120.183603} {\bibfield  {journal} {\bibinfo  {journal}
  {Phys. Rev. Lett.}\ }\textbf {\bibinfo {volume} {120}},\ \bibinfo {pages}
  {183603} (\bibinfo {year} {2018})}\BibitemShut {NoStop}%
\bibitem [{\citenamefont {Dogra}\ \emph {et~al.}(2019)\citenamefont {Dogra},
  \citenamefont {Landini}, \citenamefont {Kroeger}, \citenamefont {Hruby},
  \citenamefont {Donner},\ and\ \citenamefont {Esslinger}}]{Dogra_2019}%
  \BibitemOpen
  \bibfield  {author} {\bibinfo {author} {\bibfnamefont {N.}~\bibnamefont
  {Dogra}}, \bibinfo {author} {\bibfnamefont {M.}~\bibnamefont {Landini}},
  \bibinfo {author} {\bibfnamefont {K.}~\bibnamefont {Kroeger}}, \bibinfo
  {author} {\bibfnamefont {L.}~\bibnamefont {Hruby}}, \bibinfo {author}
  {\bibfnamefont {T.}~\bibnamefont {Donner}}, \ and\ \bibinfo {author}
  {\bibfnamefont {T.}~\bibnamefont {Esslinger}},\ }\href@noop {} {\bibfield
  {journal} {\bibinfo  {journal} {arXiv:1901.05974}\ } (\bibinfo {year}
  {2019})}\BibitemShut {NoStop}%
\bibitem [{\citenamefont {Rodr\'{\i}guez~Chiacchio}\ and\ \citenamefont
  {Nunnenkamp}(2019)}]{Chiacchio_2019}%
  \BibitemOpen
  \bibfield  {author} {\bibinfo {author} {\bibfnamefont {E.~I.}\ \bibnamefont
  {Rodr\'{\i}guez~Chiacchio}}\ and\ \bibinfo {author} {\bibfnamefont
  {A.}~\bibnamefont {Nunnenkamp}},\ }\href {\doibase
  10.1103/PhysRevLett.122.193605} {\bibfield  {journal} {\bibinfo  {journal}
  {Phys. Rev. Lett.}\ }\textbf {\bibinfo {volume} {122}},\ \bibinfo {pages}
  {193605} (\bibinfo {year} {2019})}\BibitemShut {NoStop}%
\bibitem [{\citenamefont {Buca}\ and\ \citenamefont
  {Jaksch}(2019)}]{Buca_2019}%
  \BibitemOpen
  \bibfield  {author} {\bibinfo {author} {\bibfnamefont {B.}~\bibnamefont
  {Buca}}\ and\ \bibinfo {author} {\bibfnamefont {D.}~\bibnamefont {Jaksch}},\
  }\href@noop {} {\bibfield  {journal} {\bibinfo  {journal} {arXiv:1905.12880}\
  } (\bibinfo {year} {2019})}\BibitemShut {NoStop}%
\bibitem [{\citenamefont {Fitzpatrick}\ \emph {et~al.}(2017)\citenamefont
  {Fitzpatrick}, \citenamefont {Sundaresan}, \citenamefont {Li}, \citenamefont
  {Koch},\ and\ \citenamefont {Houck}}]{Fitzpatrick_2017}%
  \BibitemOpen
  \bibfield  {author} {\bibinfo {author} {\bibfnamefont {M.}~\bibnamefont
  {Fitzpatrick}}, \bibinfo {author} {\bibfnamefont {N.~M.}\ \bibnamefont
  {Sundaresan}}, \bibinfo {author} {\bibfnamefont {A.~C.~Y.}\ \bibnamefont
  {Li}}, \bibinfo {author} {\bibfnamefont {J.}~\bibnamefont {Koch}}, \ and\
  \bibinfo {author} {\bibfnamefont {A.~A.}\ \bibnamefont {Houck}},\ }\href
  {\doibase 10.1103/PhysRevX.7.011016} {\bibfield  {journal} {\bibinfo
  {journal} {Phys. Rev. X}\ }\textbf {\bibinfo {volume} {7}},\ \bibinfo {pages}
  {011016} (\bibinfo {year} {2017})}\BibitemShut {NoStop}%
\bibitem [{\citenamefont {Fink}\ \emph {et~al.}(2018)\citenamefont {Fink},
  \citenamefont {Schade}, \citenamefont {H{\"o}fling}, \citenamefont
  {Schneider},\ and\ \citenamefont {Imamoglu}}]{Fink_2018}%
  \BibitemOpen
  \bibfield  {author} {\bibinfo {author} {\bibfnamefont {T.}~\bibnamefont
  {Fink}}, \bibinfo {author} {\bibfnamefont {A.}~\bibnamefont {Schade}},
  \bibinfo {author} {\bibfnamefont {S.}~\bibnamefont {H{\"o}fling}}, \bibinfo
  {author} {\bibfnamefont {C.}~\bibnamefont {Schneider}}, \ and\ \bibinfo
  {author} {\bibfnamefont {A.}~\bibnamefont {Imamoglu}},\ }\href {\doibase
  10.1038/s41567-017-0020-9} {\bibfield  {journal} {\bibinfo  {journal} {Nature
  Physics}\ }\textbf {\bibinfo {volume} {14}},\ \bibinfo {pages} {365}
  (\bibinfo {year} {2018})}\BibitemShut {NoStop}%
\bibitem [{\citenamefont {K\'onya}\ \emph {et~al.}(2018)\citenamefont
  {K\'onya}, \citenamefont {Nagy}, \citenamefont {Szirmai},\ and\ \citenamefont
  {Domokos}}]{Konya_2018}%
  \BibitemOpen
  \bibfield  {author} {\bibinfo {author} {\bibfnamefont {G.}~\bibnamefont
  {K\'onya}}, \bibinfo {author} {\bibfnamefont {D.}~\bibnamefont {Nagy}},
  \bibinfo {author} {\bibfnamefont {G.}~\bibnamefont {Szirmai}}, \ and\
  \bibinfo {author} {\bibfnamefont {P.}~\bibnamefont {Domokos}},\ }\href
  {\doibase 10.1103/PhysRevA.98.063608} {\bibfield  {journal} {\bibinfo
  {journal} {Phys. Rev. A}\ }\textbf {\bibinfo {volume} {98}},\ \bibinfo
  {pages} {063608} (\bibinfo {year} {2018})}\BibitemShut {NoStop}%
\bibitem [{\citenamefont {Puebla}\ \emph {et~al.}(2019)\citenamefont {Puebla},
  \citenamefont {Casanova}, \citenamefont {Houhou}, \citenamefont {Solano},\
  and\ \citenamefont {Paternostro}}]{Puebla_2019}%
  \BibitemOpen
  \bibfield  {author} {\bibinfo {author} {\bibfnamefont {R.}~\bibnamefont
  {Puebla}}, \bibinfo {author} {\bibfnamefont {J.}~\bibnamefont {Casanova}},
  \bibinfo {author} {\bibfnamefont {O.}~\bibnamefont {Houhou}}, \bibinfo
  {author} {\bibfnamefont {E.}~\bibnamefont {Solano}}, \ and\ \bibinfo {author}
  {\bibfnamefont {M.}~\bibnamefont {Paternostro}},\ }\href {\doibase
  10.1103/PhysRevA.99.032303} {\bibfield  {journal} {\bibinfo  {journal} {Phys.
  Rev. A}\ }\textbf {\bibinfo {volume} {99}},\ \bibinfo {pages} {032303}
  (\bibinfo {year} {2019})}\BibitemShut {NoStop}%
\bibitem [{\citenamefont {Zamora}\ \emph {et~al.}(2017)\citenamefont {Zamora},
  \citenamefont {Sieberer}, \citenamefont {Dunnett}, \citenamefont {Diehl},\
  and\ \citenamefont {Szyma\ifmmode~\acute{n}\else
  \'{n}\fi{}ska}}]{Zamora_2017}%
  \BibitemOpen
  \bibfield  {author} {\bibinfo {author} {\bibfnamefont {A.}~\bibnamefont
  {Zamora}}, \bibinfo {author} {\bibfnamefont {L.~M.}\ \bibnamefont
  {Sieberer}}, \bibinfo {author} {\bibfnamefont {K.}~\bibnamefont {Dunnett}},
  \bibinfo {author} {\bibfnamefont {S.}~\bibnamefont {Diehl}}, \ and\ \bibinfo
  {author} {\bibfnamefont {M.~H.}\ \bibnamefont {Szyma\ifmmode~\acute{n}\else
  \'{n}\fi{}ska}},\ }\href {\doibase 10.1103/PhysRevX.7.041006} {\bibfield
  {journal} {\bibinfo  {journal} {Phys. Rev. X}\ }\textbf {\bibinfo {volume}
  {7}},\ \bibinfo {pages} {041006} (\bibinfo {year} {2017})}\BibitemShut
  {NoStop}%
\bibitem [{\citenamefont {Zhang}\ \emph {et~al.}(2019)\citenamefont {Zhang},
  \citenamefont {Agarwal},\ and\ \citenamefont {Scully}}]{Zhang_2019}%
  \BibitemOpen
  \bibfield  {author} {\bibinfo {author} {\bibfnamefont {Z.}~\bibnamefont
  {Zhang}}, \bibinfo {author} {\bibfnamefont {G.~S.}\ \bibnamefont {Agarwal}},
  \ and\ \bibinfo {author} {\bibfnamefont {M.~O.}\ \bibnamefont {Scully}},\
  }\href {\doibase 10.1103/PhysRevLett.122.158101} {\bibfield  {journal}
  {\bibinfo  {journal} {Phys. Rev. Lett.}\ }\textbf {\bibinfo {volume} {122}},\
  \bibinfo {pages} {158101} (\bibinfo {year} {2019})}\BibitemShut {NoStop}%
\bibitem [{\citenamefont {Xu}\ and\ \citenamefont {Pu}(2019)}]{Xu_2019}%
  \BibitemOpen
  \bibfield  {author} {\bibinfo {author} {\bibfnamefont {Y.}~\bibnamefont
  {Xu}}\ and\ \bibinfo {author} {\bibfnamefont {H.}~\bibnamefont {Pu}},\ }\href
  {\doibase 10.1103/PhysRevLett.122.193201} {\bibfield  {journal} {\bibinfo
  {journal} {Phys. Rev. Lett.}\ }\textbf {\bibinfo {volume} {122}},\ \bibinfo
  {pages} {193201} (\bibinfo {year} {2019})}\BibitemShut {NoStop}%
\bibitem [{\citenamefont {Heiss}(2012)}]{Heiss_2012}%
  \BibitemOpen
  \bibfield  {author} {\bibinfo {author} {\bibfnamefont {W.~D.}\ \bibnamefont
  {Heiss}},\ }\href {\doibase 10.1088/1751-8113/45/44/444016} {\bibfield
  {journal} {\bibinfo  {journal} {Journal of Physics A: Mathematical and
  Theoretical}\ }\textbf {\bibinfo {volume} {45}},\ \bibinfo {pages} {444016}
  (\bibinfo {year} {2012})}\BibitemShut {NoStop}%
\bibitem [{\citenamefont {{\"O}zdemir}\ \emph {et~al.}(2019)\citenamefont
  {{\"O}zdemir}, \citenamefont {Rotter}, \citenamefont {Nori},\ and\
  \citenamefont {Yang}}]{Ozdemir_2019}%
  \BibitemOpen
  \bibfield  {author} {\bibinfo {author} {\bibfnamefont {{\c S}.~K.}\
  \bibnamefont {{\"O}zdemir}}, \bibinfo {author} {\bibfnamefont
  {S.}~\bibnamefont {Rotter}}, \bibinfo {author} {\bibfnamefont
  {F.}~\bibnamefont {Nori}}, \ and\ \bibinfo {author} {\bibfnamefont
  {L.}~\bibnamefont {Yang}},\ }\href {\doibase 10.1038/s41563-019-0304-9}
  {\bibfield  {journal} {\bibinfo  {journal} {Nature Materials}\ }\textbf
  {\bibinfo {volume} {18}},\ \bibinfo {pages} {783} (\bibinfo {year}
  {2019})}\BibitemShut {NoStop}%
\bibitem [{\citenamefont {Hanai}\ \emph {et~al.}(2019)\citenamefont {Hanai},
  \citenamefont {Edelman}, \citenamefont {Ohashi},\ and\ \citenamefont
  {Littlewood}}]{Hanai_2019}%
  \BibitemOpen
  \bibfield  {author} {\bibinfo {author} {\bibfnamefont {R.}~\bibnamefont
  {Hanai}}, \bibinfo {author} {\bibfnamefont {A.}~\bibnamefont {Edelman}},
  \bibinfo {author} {\bibfnamefont {Y.}~\bibnamefont {Ohashi}}, \ and\ \bibinfo
  {author} {\bibfnamefont {P.~B.}\ \bibnamefont {Littlewood}},\ }\href
  {\doibase 10.1103/PhysRevLett.122.185301} {\bibfield  {journal} {\bibinfo
  {journal} {Phys. Rev. Lett.}\ }\textbf {\bibinfo {volume} {122}},\ \bibinfo
  {pages} {185301} (\bibinfo {year} {2019})}\BibitemShut {NoStop}%
\bibitem [{\citenamefont {Bhaseen}\ \emph {et~al.}(2012)\citenamefont
  {Bhaseen}, \citenamefont {Mayoh}, \citenamefont {Simons},\ and\ \citenamefont
  {Keeling}}]{Bhaseen_2012}%
  \BibitemOpen
  \bibfield  {author} {\bibinfo {author} {\bibfnamefont {M.~J.}\ \bibnamefont
  {Bhaseen}}, \bibinfo {author} {\bibfnamefont {J.}~\bibnamefont {Mayoh}},
  \bibinfo {author} {\bibfnamefont {B.~D.}\ \bibnamefont {Simons}}, \ and\
  \bibinfo {author} {\bibfnamefont {J.}~\bibnamefont {Keeling}},\ }\href
  {\doibase 10.1103/PhysRevA.85.013817} {\bibfield  {journal} {\bibinfo
  {journal} {Phys. Rev. A}\ }\textbf {\bibinfo {volume} {85}},\ \bibinfo
  {pages} {013817} (\bibinfo {year} {2012})}\BibitemShut {NoStop}%
\bibitem [{\citenamefont {Baksic}\ and\ \citenamefont
  {Ciuti}(2014)}]{Baksic_2014}%
  \BibitemOpen
  \bibfield  {author} {\bibinfo {author} {\bibfnamefont {A.}~\bibnamefont
  {Baksic}}\ and\ \bibinfo {author} {\bibfnamefont {C.}~\bibnamefont {Ciuti}},\
  }\href {\doibase 10.1103/PhysRevLett.112.173601} {\bibfield  {journal}
  {\bibinfo  {journal} {Phys. Rev. Lett.}\ }\textbf {\bibinfo {volume} {112}},\
  \bibinfo {pages} {173601} (\bibinfo {year} {2014})}\BibitemShut {NoStop}%
\bibitem [{\citenamefont {Kirton}\ \emph {et~al.}(2019)\citenamefont {Kirton},
  \citenamefont {Roses}, \citenamefont {Keeling},\ and\ \citenamefont
  {Dalla~Torre}}]{Kirton_2019}%
  \BibitemOpen
  \bibfield  {author} {\bibinfo {author} {\bibfnamefont {P.}~\bibnamefont
  {Kirton}}, \bibinfo {author} {\bibfnamefont {M.~M.}\ \bibnamefont {Roses}},
  \bibinfo {author} {\bibfnamefont {J.}~\bibnamefont {Keeling}}, \ and\
  \bibinfo {author} {\bibfnamefont {E.~G.}\ \bibnamefont {Dalla~Torre}},\
  }\href {\doibase doi:10.1002/qute.201800043} {\bibfield  {journal} {\bibinfo
  {journal} {Advanced Quantum Technologies}\ }\textbf {\bibinfo {volume} {2}},\
  \bibinfo {pages} {1800043} (\bibinfo {year} {2019})}\BibitemShut {NoStop}%
\bibitem [{\citenamefont {Joe}\ \emph {et~al.}(2006)\citenamefont {Joe},
  \citenamefont {Satanin},\ and\ \citenamefont {Kim}}]{Joe_2006}%
  \BibitemOpen
  \bibfield  {author} {\bibinfo {author} {\bibfnamefont {Y.~S.}\ \bibnamefont
  {Joe}}, \bibinfo {author} {\bibfnamefont {A.~M.}\ \bibnamefont {Satanin}}, \
  and\ \bibinfo {author} {\bibfnamefont {C.~S.}\ \bibnamefont {Kim}},\ }\href
  {\doibase 10.1088/0031-8949/74/2/020} {\bibfield  {journal} {\bibinfo
  {journal} {Physica Scripta}\ }\textbf {\bibinfo {volume} {74}},\ \bibinfo
  {pages} {259} (\bibinfo {year} {2006})}\BibitemShut {NoStop}%
\bibitem [{\citenamefont {Kamenev}(2011)}]{kamenev}%
  \BibitemOpen
  \bibfield  {author} {\bibinfo {author} {\bibfnamefont {A.}~\bibnamefont
  {Kamenev}},\ }\href@noop {} {\emph {\bibinfo {title} {Field theory of
  non-equilibrium systems}}}\ (\bibinfo  {publisher} {Cambridge University
  Press},\ \bibinfo {year} {2011})\BibitemShut {NoStop}%
\bibitem [{\citenamefont {Dalla~Torre}\ \emph {et~al.}(2013)\citenamefont
  {Dalla~Torre}, \citenamefont {Diehl}, \citenamefont {Lukin}, \citenamefont
  {Sachdev},\ and\ \citenamefont {Strack}}]{DallaTorre_2013}%
  \BibitemOpen
  \bibfield  {author} {\bibinfo {author} {\bibfnamefont {E.~G.}\ \bibnamefont
  {Dalla~Torre}}, \bibinfo {author} {\bibfnamefont {S.}~\bibnamefont {Diehl}},
  \bibinfo {author} {\bibfnamefont {M.~D.}\ \bibnamefont {Lukin}}, \bibinfo
  {author} {\bibfnamefont {S.}~\bibnamefont {Sachdev}}, \ and\ \bibinfo
  {author} {\bibfnamefont {P.}~\bibnamefont {Strack}},\ }\href {\doibase
  10.1103/physreva.87.023831} {\bibfield  {journal} {\bibinfo  {journal}
  {Physical Review A}\ }\textbf {\bibinfo {volume} {87}},\ \bibinfo {pages}
  {023831} (\bibinfo {year} {2013})}\BibitemShut {NoStop}%
\bibitem [{\citenamefont {Sieberer}\ \emph {et~al.}(2016)\citenamefont
  {Sieberer}, \citenamefont {Buchhold},\ and\ \citenamefont
  {Diehl}}]{Sieberer_2016}%
  \BibitemOpen
  \bibfield  {author} {\bibinfo {author} {\bibfnamefont {L.~M.}\ \bibnamefont
  {Sieberer}}, \bibinfo {author} {\bibfnamefont {M.}~\bibnamefont {Buchhold}},
  \ and\ \bibinfo {author} {\bibfnamefont {S.}~\bibnamefont {Diehl}},\ }\href
  {\doibase 10.1088/0034-4885/79/9/096001} {\bibfield  {journal} {\bibinfo
  {journal} {Reports on Progress in Physics}\ }\textbf {\bibinfo {volume}
  {79}},\ \bibinfo {pages} {096001} (\bibinfo {year} {2016})}\BibitemShut
  {NoStop}%
\bibitem [{\citenamefont {Bogoliubov}(1947)}]{Bogolyubov_1947}%
  \BibitemOpen
  \bibfield  {author} {\bibinfo {author} {\bibfnamefont {N.}~\bibnamefont
  {Bogoliubov}},\ }\href@noop {} {\bibfield  {journal} {\bibinfo  {journal} {J.
  Phys}\ }\textbf {\bibinfo {volume} {11}},\ \bibinfo {pages} {23} (\bibinfo
  {year} {1947})}\BibitemShut {NoStop}%
\bibitem [{\citenamefont {Valatin}(1958)}]{Valatin_1958}%
  \BibitemOpen
  \bibfield  {author} {\bibinfo {author} {\bibfnamefont {J.}~\bibnamefont
  {Valatin}},\ }\href@noop {} {\bibfield  {journal} {\bibinfo  {journal} {Il
  Nuovo Cimento (1955-1965)}\ }\textbf {\bibinfo {volume} {7}},\ \bibinfo
  {pages} {843} (\bibinfo {year} {1958})}\BibitemShut {NoStop}%
\bibitem [{\citenamefont {Xiao}(2009)}]{Xiao_2009}%
  \BibitemOpen
  \bibfield  {author} {\bibinfo {author} {\bibfnamefont {M.-w.}\ \bibnamefont
  {Xiao}},\ }\href@noop {} {\bibfield  {journal} {\bibinfo  {journal}
  {arXiv:0908.0787}\ } (\bibinfo {year} {2009})}\BibitemShut {NoStop}%
\bibitem [{\citenamefont {Prosen}(2008)}]{Prosen_2008}%
  \BibitemOpen
  \bibfield  {author} {\bibinfo {author} {\bibfnamefont {T.}~\bibnamefont
  {Prosen}},\ }\href {\doibase 10.1088/1367-2630/10/4/043026} {\bibfield
  {journal} {\bibinfo  {journal} {New Journal of Physics}\ }\textbf {\bibinfo
  {volume} {10}},\ \bibinfo {pages} {043026} (\bibinfo {year}
  {2008})}\BibitemShut {NoStop}%
\bibitem [{\citenamefont {Prosen}\ and\ \citenamefont
  {Seligman}(2010)}]{Prosen_2010}%
  \BibitemOpen
  \bibfield  {author} {\bibinfo {author} {\bibfnamefont {T.}~\bibnamefont
  {Prosen}}\ and\ \bibinfo {author} {\bibfnamefont {T.~H.}\ \bibnamefont
  {Seligman}},\ }\href {\doibase 10.1088/1751-8113/43/39/392004} {\bibfield
  {journal} {\bibinfo  {journal} {Journal of Physics A: Mathematical and
  Theoretical}\ }\textbf {\bibinfo {volume} {43}},\ \bibinfo {pages} {392004}
  (\bibinfo {year} {2010})}\BibitemShut {NoStop}%
\bibitem [{\citenamefont {Kohler}\ \emph {et~al.}(2018)\citenamefont {Kohler},
  \citenamefont {Gerber}, \citenamefont {Dowd},\ and\ \citenamefont
  {Stamper-Kurn}}]{Kohler_2018}%
  \BibitemOpen
  \bibfield  {author} {\bibinfo {author} {\bibfnamefont {J.}~\bibnamefont
  {Kohler}}, \bibinfo {author} {\bibfnamefont {J.~A.}\ \bibnamefont {Gerber}},
  \bibinfo {author} {\bibfnamefont {E.}~\bibnamefont {Dowd}}, \ and\ \bibinfo
  {author} {\bibfnamefont {D.~M.}\ \bibnamefont {Stamper-Kurn}},\ }\href
  {\doibase 10.1103/PhysRevLett.120.013601} {\bibfield  {journal} {\bibinfo
  {journal} {Phys. Rev. Lett.}\ }\textbf {\bibinfo {volume} {120}},\ \bibinfo
  {pages} {013601} (\bibinfo {year} {2018})}\BibitemShut {NoStop}%
\bibitem [{\citenamefont {Dicke}(1954)}]{Dicke_1954}%
  \BibitemOpen
  \bibfield  {author} {\bibinfo {author} {\bibfnamefont {R.~H.}\ \bibnamefont
  {Dicke}},\ }\href {\doibase 10.1103/PhysRev.93.99} {\bibfield  {journal}
  {\bibinfo  {journal} {Phys. Rev.}\ }\textbf {\bibinfo {volume} {93}},\
  \bibinfo {pages} {99} (\bibinfo {year} {1954})}\BibitemShut {NoStop}%
\bibitem [{\citenamefont {Tavis}\ and\ \citenamefont
  {Cummings}(1968)}]{Tavis_1968}%
  \BibitemOpen
  \bibfield  {author} {\bibinfo {author} {\bibfnamefont {M.}~\bibnamefont
  {Tavis}}\ and\ \bibinfo {author} {\bibfnamefont {F.~W.}\ \bibnamefont
  {Cummings}},\ }\href {\doibase 10.1103/PhysRev.170.379} {\bibfield  {journal}
  {\bibinfo  {journal} {Phys. Rev.}\ }\textbf {\bibinfo {volume} {170}},\
  \bibinfo {pages} {379} (\bibinfo {year} {1968})}\BibitemShut {NoStop}%
\bibitem [{\citenamefont {Kirton}\ and\ \citenamefont
  {Keeling}(2017)}]{Kirton_2017}%
  \BibitemOpen
  \bibfield  {author} {\bibinfo {author} {\bibfnamefont {P.}~\bibnamefont
  {Kirton}}\ and\ \bibinfo {author} {\bibfnamefont {J.}~\bibnamefont
  {Keeling}},\ }\href {\doibase 10.1103/PhysRevLett.118.123602} {\bibfield
  {journal} {\bibinfo  {journal} {Phys. Rev. Lett.}\ }\textbf {\bibinfo
  {volume} {118}},\ \bibinfo {pages} {123602} (\bibinfo {year}
  {2017})}\BibitemShut {NoStop}%
\bibitem [{\citenamefont {Emary}\ and\ \citenamefont
  {Brandes}(2003)}]{Emary_2003}%
  \BibitemOpen
  \bibfield  {author} {\bibinfo {author} {\bibfnamefont {C.}~\bibnamefont
  {Emary}}\ and\ \bibinfo {author} {\bibfnamefont {T.}~\bibnamefont
  {Brandes}},\ }\href {\doibase 10.1103/PhysRevE.67.066203} {\bibfield
  {journal} {\bibinfo  {journal} {Phys. Rev. E}\ }\textbf {\bibinfo {volume}
  {67}},\ \bibinfo {pages} {066203} (\bibinfo {year} {2003})}\BibitemShut
  {NoStop}%
\bibitem [{\citenamefont {Dimer}\ \emph {et~al.}(2007)\citenamefont {Dimer},
  \citenamefont {Estienne}, \citenamefont {Parkins},\ and\ \citenamefont
  {Carmichael}}]{Dimer_2007}%
  \BibitemOpen
  \bibfield  {author} {\bibinfo {author} {\bibfnamefont {F.}~\bibnamefont
  {Dimer}}, \bibinfo {author} {\bibfnamefont {B.}~\bibnamefont {Estienne}},
  \bibinfo {author} {\bibfnamefont {A.~S.}\ \bibnamefont {Parkins}}, \ and\
  \bibinfo {author} {\bibfnamefont {H.~J.}\ \bibnamefont {Carmichael}},\ }\href
  {\doibase 10.1103/PhysRevA.75.013804} {\bibfield  {journal} {\bibinfo
  {journal} {Phys. Rev. A}\ }\textbf {\bibinfo {volume} {75}},\ \bibinfo
  {pages} {013804} (\bibinfo {year} {2007})}\BibitemShut {NoStop}%
\bibitem [{\citenamefont {Fano}(1961)}]{Fano_1961}%
  \BibitemOpen
  \bibfield  {author} {\bibinfo {author} {\bibfnamefont {U.}~\bibnamefont
  {Fano}},\ }\href {\doibase 10.1103/PhysRev.124.1866} {\bibfield  {journal}
  {\bibinfo  {journal} {Phys. Rev.}\ }\textbf {\bibinfo {volume} {124}},\
  \bibinfo {pages} {1866} (\bibinfo {year} {1961})}\BibitemShut {NoStop}%
\bibitem [{im_()}]{im_freq}%
  \BibitemOpen
  \href@noop {} {}\bibinfo {note} {In the closed system, stable phases are
  characterized by real eigenfrequencies.}\BibitemShut {Stop}%
\bibitem [{\citenamefont {Scarlatella}\ \emph {et~al.}(2019)\citenamefont
  {Scarlatella}, \citenamefont {Clerk},\ and\ \citenamefont
  {Schiro}}]{Scarlatella_2019}%
  \BibitemOpen
  \bibfield  {author} {\bibinfo {author} {\bibfnamefont {O.}~\bibnamefont
  {Scarlatella}}, \bibinfo {author} {\bibfnamefont {A.~A.}\ \bibnamefont
  {Clerk}}, \ and\ \bibinfo {author} {\bibfnamefont {M.}~\bibnamefont
  {Schiro}},\ }\href@noop {} {\bibfield  {journal} {\bibinfo  {journal} {New
  Journal of Physics}\ }\textbf {\bibinfo {volume} {21}},\ \bibinfo {pages}
  {043040} (\bibinfo {year} {2019})}\BibitemShut {NoStop}%
\bibitem [{\citenamefont {Scully}\ and\ \citenamefont
  {Zubairy}(1999)}]{Scully}%
  \BibitemOpen
  \bibfield  {author} {\bibinfo {author} {\bibfnamefont {M.~O.}\ \bibnamefont
  {Scully}}\ and\ \bibinfo {author} {\bibfnamefont {M.~S.}\ \bibnamefont
  {Zubairy}},\ }\href@noop {} {\emph {\bibinfo {title} {Quantum optics}}}\
  (\bibinfo  {publisher} {AAPT},\ \bibinfo {year} {1999})\BibitemShut {NoStop}%
\end{thebibliography}

%merlin.mbs apsrev4-1.bst 2010-07-25 4.21a (PWD, AO, DPC) hacked
%Control: key (0)
%Control: author (8) initials jnrlst
%Control: editor formatted (1) identically to author
%Control: production of article title (-1) disabled
%Control: page (0) single
%Control: year (1) truncated
%Control: production of eprint (0) enabled
%

\end{document}